\documentclass[format=acmsmall, review=false, screen=true, numbers=true,nonacm=true]{acmart}

\pdfoutput=1

\usepackage{booktabs} %
\usepackage{multirow}
\usepackage{array}
\newcolumntype{$}{>{\global\let\currentrowstyle\relax}}
\newcolumntype{^}{>{\currentrowstyle}}
\newcommand{\rowstyle}[1]{\gdef\currentrowstyle{#1}%
	#1\ignorespaces
}
\usepackage{enumitem}
\newcommand{\citeAut}[1]{\begin{NoHyper}{\bf\citeauthor{#1}}\end{NoHyper}
}
\newcommand{\citeAutRef}[1]{\citeAut{#1}~\cite{#1}}

\usepackage[ruled]{algorithm2e} %

\newtheorem{mydef2}{Definition}[section]{\bfseries}{\itshape}
\SetAlFnt{\small}
\SetAlCapFnt{\small}
\SetAlCapNameFnt{\small}
\SetAlCapHSkip{0pt}
\IncMargin{-\parindent}
\usepackage{amssymb}%
\usepackage{pifont}%
\newcommand{\cmark}{\ding{51}}%
\newcommand{\xmark}{\ding{55}}%
\acmJournal{CSUR}
\acmVolume{1}
\acmNumber{1}
\acmYear{2019}
\acmMonth{10}
\copyrightyear{2019}
\acmDOI{0000001.0000001}
\definecolor{pcolor}{rgb}{0,0.6,0}
\begin{document}
 \title[A Survey on Subgraph Counting]{A Survey on Subgraph Counting: Concepts, Algorithms \\ and Applications to Network Motifs and Graphlets}
\author{Pedro Ribeiro$^{1}$}
\author{Pedro Paredes$^{1,2}$}
\author{Miguel E.P. Silva$^{1,3}$}
\author{David Apar\'icio$^{1,4}$}
\author{Fernando Silva$^{1}$}

\affiliation{\vspace{0.1cm} \\ $^{1}$ \institution{DCC-FCUP \& CRACS/INESC-TEC, University of Porto}\country{Portugal} \\
$^{2}$ \institution{Carnegie Mellon University}\country{USA} \\
$^{3}$ \institution{University of Manchester}\country{UK} \\
$^{4}$ \institution{Feedzai}
}

\begin{abstract}

Computing subgraph frequencies is a fundamental task that lies at the core of several network analysis methodologies, such as network motifs and graphlet-based metrics, which have been widely used to categorize and compare networks from multiple domains. Counting subgraphs is however computationally very expensive and there has been a large body of work on efficient algorithms and strategies to make subgraph counting feasible for larger subgraphs and networks.

This survey aims precisely to provide a comprehensive overview of the existing methods for subgraph counting. Our main contribution is a general and structured review of existing algorithms, classifying them on a set of key characteristics, highlighting their main similarities and differences. We identify and describe the main conceptual approaches, giving insight on their advantages and limitations, and provide pointers to existing implementations. We initially focus on exact sequential algorithms, but we also do a thorough survey on approximate methodologies (with a trade-off between accuracy and execution time) and parallel strategies (that need to deal with an unbalanced search space).

\end{abstract}

\maketitle

\vspace{-0.1cm}
\normalsize

\renewcommand{\shortauthors}{P. Ribeiro, P. Paredes, M. E. P. Silva, D. Apar\'icio, and F. Silva}

\section{Introduction}

Networks (or graphs) are a very flexible and powerful way of modeling many real-world systems. In its essence, they capture the interactions of a system, by representing entities as nodes and their relations as edges connecting them (e.g., people are nodes in social networks and edges connect those that have some relationship between them, such as friendships or citations). Networks have thus been used to analyze all kinds of social, biological and communication processes~\cite{costa2011analyzing}. Extracting information from networks is therefore a vital interdisciplinary task that has been emerging as a research area by itself, commonly known as Network Science~\cite{lewis2011network,barabasi2016network}.

One very common and important methodology is to look at the networks from a subgraph perspective, identifying the characteristic and recurrent connection patterns. For instance, network motif analysis~\cite{milo2002network} has identified the feed-forward loop as a recurring and crucial functional pattern in many real biological networks, such as gene regulation and metabolic networks \cite{mangan2003structure, zhu2005structural}. Another example is the usage of graphlet-degree distributions to show that protein-protein interaction networks are more akin to geometric graphs than with traditional scale-free models~\cite{przulj2007}.

At the heart of these topologically rich approaches lies the subgraph counting problem, that is, the ability to compute subgraph frequencies. However, this is a very hard computational task. In fact, determining if one subgraph exists at all in another larger network (i.e., \textit{subgraph isomorphism}~\cite{ullmann1976algorithm}) is an \mbox{NP-Complete} problem~\cite{cook1971complexity}. Determining the exact frequency is even harder, and millions or even billions of subgraph occurrences are typically found even in relatively small networks.

Given both its usefulness and hard tractability, subgraph counting has been raising a considerable amount of interest from the research community, with a large body of published literature. This survey aims precisely to organize and summarize these research results, providing a comprehensive overview of the field. Our main contributions are the following:

\begin{itemize}
	\item \textbf{A comprehensive review of algorithms for \textit{exact} subgraph counting.} We give a structured historical perspective on algorithms for computing exact subgraph frequencies. We provide a complete overview table in which we employ a taxonomy that allows to classify all algorithms on a set of key characteristics, highlighting their main similarities and differences. We also identify and describe the main conceptual ideas, giving insight on their main advantages and possible limitations. We also provide links to existing implementations, exposing which approaches are readily available.
		
	\item \textbf{A comprehensive review of algorithms for \textit{approximate} subgraph counting.} Given the hardness of the problem, many authors have resorted to approximation schemes, which allow trading some accuracy for faster execution times. As on the exact case, we provide historical context, links to implementations and we give a classification and description of key properties, explaining how the existing approaches deal with the balance between precision and running time.
	
	\item \textbf{A comprehensive review of \textit{parallel} subgraph counting methodologies.} It is only natural that researchers have tried to harness the power of parallel architectures to provide scalable approaches that might decrease the needed computation time. As before, we provide an historical overview,  coupled with classification on a set of important aspects, such as the type of parallel platform or availability of an implementation. We also give particular attention to how the methodologies tackle the unbalanced nature of the search space.
\end{itemize} 

We complement this journey trough the algorithmic strategies with a \textit{clear formal definition of the subgraph counting problem} being discussed here, an \textit{overview of its applications} and complete and a large number of \textit{references to related work} that is not directly in the scope of this article. We believe that this survey provides the reader with an insightful and complete perspective on the field, both from a methodological and an application point of view.

The remainder of this paper is structured as follows. Section~\ref{sec:preliminaries} presents necessary terminology, formally describes subgraph counting, and describes possible applications related subgraph counting. Section~\ref{sec:exact} reviews exact algorithms, divided between full enumeration and analytical methods. Approximate algorithms are described in Section~\ref{sec:sampling} and parallel strategies are presented in Section~\ref{sec:parallel}. Finally, in Section~\ref{sec:conclusions} we give our concluding remarks.

\newpage

\section{Preliminaries}\label{sec:preliminaries}

\subsection{Concepts and Common Terminology}
\label{sec:terminology}

This section introduces concepts and terminology related to subgraph counting that will be used throughout this paper. A \emph{network} is modeled with the mathematical object \emph{graph}, and the two terms are used interchangeably. Networks considered in this work are \emph{simple labeled graphs}. Here we are interested in algorithms that count \emph{small, connected, non-isomorphic subgraphs} on a single network.

\smallskip

\begin{itemize}[label={},leftmargin=*,itemsep=7pt]
	\item \textbf{Graph:} A graph $G$ is comprised of a set $V(G)$ of \emph{vertices/nodes} and a set $E(G)$ of \emph{edges/connections}. Nodes represent entities and edges correspond to relationships between them. Edges are represented as pairs of vertices of the form $(u, v)$, where $u, v \in V(G)$. In \emph{directed} graphs, edges $(u, v)$ are ordered pairs ($u \rightarrow v$) whereas in \emph{undirected} graphs there is no order since nodes are always reciprocally connected ($u \rightleftarrows v$). The \emph{size} of a graph is the number of vertices in the graph and it is written as $|V(G)|$. A $k$-graph is a graph of size $k$. A graph is considered \emph{simple} if it does not contain multiple edges (two or more edges connecting the same vertex pair) nor self-loops (an edge connecting a vertex to itself). Nodes are \emph{labeled} from 0 to $|V(G)|-1$, and $L(u) < L(v)$ means than $u$ has a smaller label than $v$.
	
	\item \textbf{Neighborhood and Degree:} The \emph{neighborhood} of vertex $u \in V(G)$, denoted as $N(u)$, is composed by the set of vertices $v \in V(G)$ such that $(u,v) \in E(G)$. The \emph{degree} of $u$, written as $deg(u)$, is the given by $|N(u)|$. The exclusive neighborhood $N_{exc}(u, S)$ are the neighbors of $u$ that are not neighbors of any $v \in S$ with $u \neq v$.
	
	\item \textbf{Graph Isomorphism:} A \emph{mapping} of a graph is a bijection where each vertex is assigned a value. In the context of this work, since graphs are \emph{labeled}, a mapping is a permutation of the node labels. Two graphs $G$ and $H$ are said to be isomorphic if there is a one-to-one mapping between the vertices of both graphs, such that there is an edge between two vertices of $G$ if and only if their corresponding vertices in $H$ also form an edge (preserving direction in the case of directed graphs). More informally, isomorphism captures the notion of two networks having the same edge structure -- the same topology -- if we ignore distinction between individual nodes. Figure~\ref{fig:isomorphic} illustrates this concept. Despite looking different, the structure of the graphs is the same, and they are isomorphic. The labels in the nodes illustrate mappings that would satisfy the conditions given for isomorphism.
	
	\begin{figure}[h]
		\vspace{-0.1cm}	
		\centering
			\includegraphics[width=0.65\linewidth]{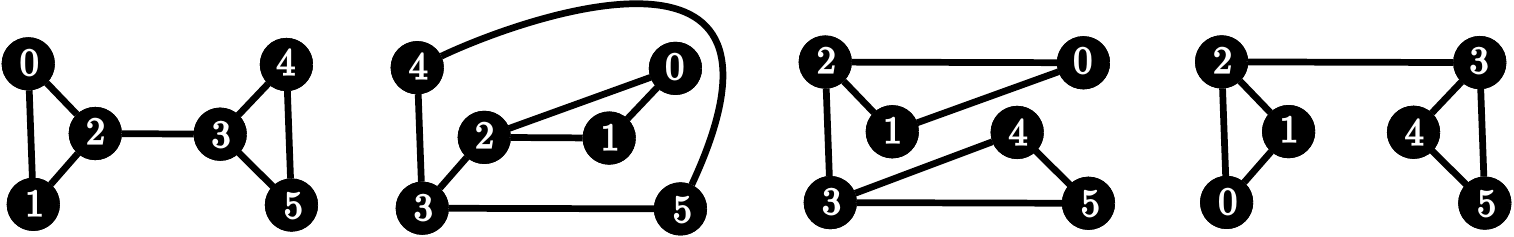}
			\vspace{-0.2cm}	
			\caption{Four isomorphic undirected graphs of size 6.}
		\label{fig:isomorphic}
		\vspace{-0.4cm}
	\end{figure}
	
	\item \textbf{Subgraphs:} A \emph{subgraph} $G_k$ of a graph $G$ is a $k$-graph such that $V(G_k) \subseteq V(G) $ and $E(G_k) \subseteq E(G)$. A subgraph is \emph{induced} if
	$\forall(u,v) \in E(G_K) \leftrightarrow (u,v) \in E(G)$ and is said to be \emph{connected} when all pairs of vertices have a sequence of edges connecting them.  \textit{Graphlets}~\cite{przulj2007} are small, connected, non-isomorphic, induced 
	subgraphs. Figure~\ref{fig:graphlets_u4} presents all 4-node undirected graphlets. 
	
	\begin{figure}[h]
	\centering
		\includegraphics[width=0.7\linewidth]{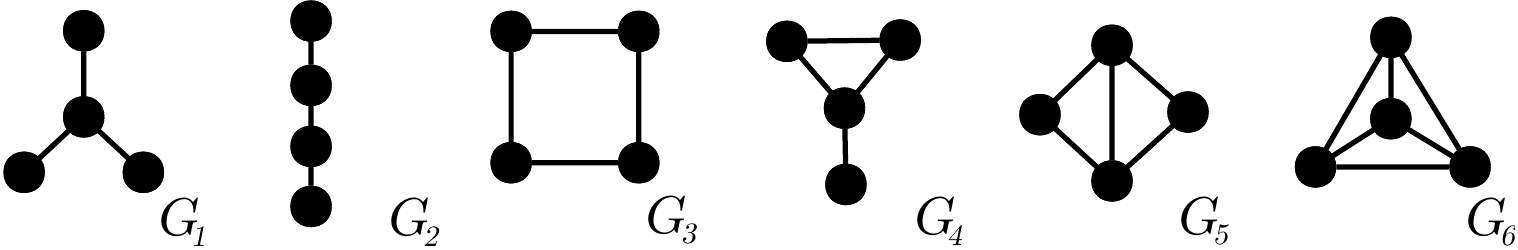}
		\vspace{-0.2cm}
		\caption{All non-isomorphic undirected subgraphs (or graphlets) of size 4.}
	\label{fig:graphlets_u4}
	\end{figure}
	
	\item \textbf{Orbit: } The set of isomorphisms of a graph into itself is called the group of \emph{automorphisms}: two vertices are said to be equivalent when there exists
	some automorphism that maps one vertex into the other. This equivalence relation
	partitions $V(G)$ into equivalence classes, which we refer to as \emph{orbits}. Therefore, \emph{orbits} are all the unique positions of a subgraph. For instance, a $k-$hub has $k$ nodes but only 2 orbits: one \emph{center-orbit} inhabited by a single node and a \emph{leaf-orbit} where the remaining $k$-1 nodes are. Nodes at the same orbit are topologically equivalent. Figure~\ref{fig:orbits_u4} shows all different orbits of the graphlets from Figure~\ref{fig:graphlets_u4}.
	
	\begin{figure}[h]
		\vspace{-0.1cm}
		\centering
		\includegraphics[width=0.65\linewidth]{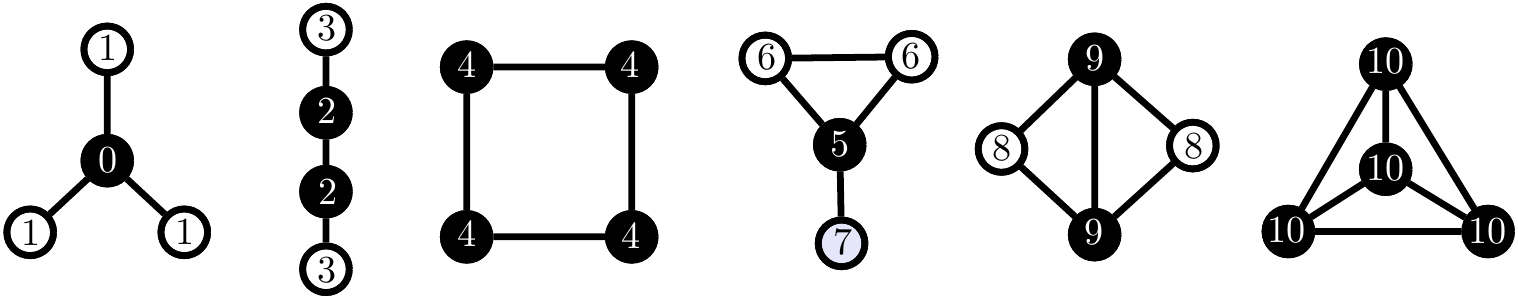}
		\vspace{-0.3cm}
		\caption{The 10 orbits of all 4-node undirected graphlets.}
		\label{fig:orbits_u4}
		\vspace{-0.4cm}
	\end{figure}

	\item \textbf{Match and Frequency:} A \emph{match} of graph $H$ in graph $G$ occurs when there is a set of nodes from $V(G)$ that induce $H$. In other words, $G_k$ is a subgraph of $G$ that is isomorphic to $H$. Figure~\ref{fig:orbits_occ} shows the matches of three different subgraphs ($A$, $B$ and $C$) on graph $G$. The \emph{frequency} of $H$ in $G$ is the number of \emph{different} $G_k \subseteq G$ that induce $H$. Two matches are considered different if they do not share all nodes and edges.

	\item \textbf{Graphlet-Degree Distribution :} It is an extension of the node-degree distribution and both can be used for graph characterization and comparison. Notice that the node-degree can be seen as simply the orbit $a$ in Figure~\ref{fig:orbits_occ}. The graphlet-degree vector $GDV(v)$ is a feature vector of $v$ specifying how many times it occurs in each orbit. The graphlet-degree distribution $GDD_G$ is a feature matrix of graph $G$ where cell $(i,j)$ indicates the number of nodes that appear $i$ times in orbit $j$, and can be constructed from $Fr_G$, the frequency matrix where each line is the $GDV$ of a single node.
	
	\begin{figure}[h]
		\centering
		\includegraphics[width=0.7\linewidth]{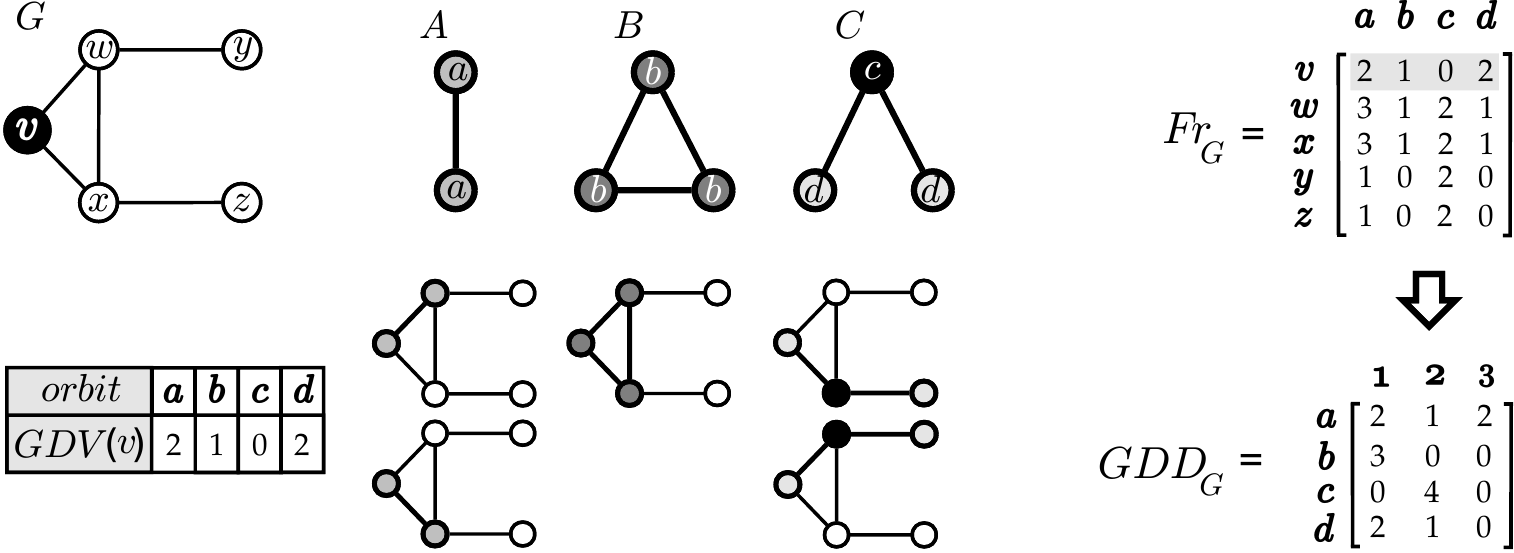}
		\caption{$GDV(v)$ obtained by enumerating all undirected graphlet-orbits of sizes 2 and 3 ($A$, $B$ and $C$) touching $v$, and resulting $Fr_G$ and $GDD_G$ matrices for the complete $3$-subgraph census
		}
		\label{fig:orbits_occ}
	\end{figure}
\end{itemize}

\subsection{Problem statement}

Making use of previous concepts and terminology, we now give a more formal definition of the problem tackled by this survey:

\begin{mydef2}[\textbf{Subgraph Counting}]
	\label{def:genproblem}
Given a set $\mathcal{G}$ of non-isomorphic subgraphs and a graph $G$, determine the frequency of all induced matches of the subgraphs $G_s \in \mathcal{G}$ in $G$. Two occurrences are considered different if they have at least one node or edge that they do not share.
\end{mydef2}

This problem is also known as \textit{subgraph census}. In short, one wants to extract the occurrences of all subgraphs of a given size, or just a smaller set of "interesting" subgraphs, contained in a large graph $G$. Note how here the input is a single graph, in contrast with Frequent Subgraph Mining (FSM) where collections of graphs are more commonly used (differences between Subgraph Counting and FSM are discussed in Section~\ref{sec:fsm}). 

Approaches diverge on which subgraphs are counted in $G$. \textit{Network-centric} methods extract all $k-$node occurrences in $G$ and then assess each occurrence's isomorphic type. On the other end of the spectrum, \textit{subgraph-centric} methods first pick a isomorphic class and then only count occurrences matching that class in $G$. Therefore, subgraph-centric methods are preferable to network-centric algorithms when only one or a few different subgraphs are to be counted. \textit{Set-centric} approaches are middle-ground algorithms that take as input a set of interesting subgraphs and only count those on $G$. This work is mainly focused on network-centric algorithms, while not limited to them, since: (a) exploring all subgraphs offers the most information possible when applying subgraph counting to a real dataset, (b) hand-picking a set of interesting subgraphs might might be hard or impossible and could be heavily dependent on our knowledge of the dataset, (c) it is intrinsically the most general approach. It is obviously possible to use subgraph-centric methods to count all isomorphic classes, simply by executing the method once per isomorphic type. However, that option is only feasible for small subgraph sizes because larger $k$ values produce too many subgraphs (see Table~\ref{tab:n_subgraphs}) and it is likely that a network only has a small subset of them, meaning that the method would spend a considerable amount of time looking for features that do not exist, while network-centric methods always do useful work since they count occurrences in the network. 

\begin{table*}[!h]
	\centering
	\small
	\caption{Number of different undirected and directed subgraphs (i.e., isomorphic classes), as well as their respective orbits, depending on the size of the graphlets.}
	\label{tab:n_subgraphs} 
	\begin{tabular}{|c||c|c||c|c|c|}
		\cline{2-5}
		
		\multicolumn{1}{c|}{}& \multicolumn{2}{c||}{Undirected} & \multicolumn{2}{c|}{Directed}     \\  \hline
		$k$       & \#Subgraphs & \#Orbits     & \#Subgraphs & \#Orbits \\  \hline  \hline
		2         & 1         & 1              & 2         & 3  (1.5  $\times$ \#Subgraphs)    \\  \hline
		3        & 2         & 3             & 15        & 30  (2.0  $\times$ \#Subgraphs)   \\  \hline
		4      & 6         & 11         & 214       & 697  (3.3  $\times$ \#Subgraphs)  \\  \hline
		5    & 21        & 58        & 9,578      & 44,907  (4.7  $\times$ \#Subgraphs)\\  \hline
		6  & 112       & 407      &    1,540,421       &   9,076,020  (5.9  $\times$ \#Subgraphs)   \\  \hline
		7  & 823       & 4,306              &      872,889,906     &    $\approx$ 7 $\times$ \#Subgraphs    \\  \hline
		8  & 11,117     & 72,489             &        1,792,473,955,306   &  $\approx$ 8 $\times$ \#Subgraphs    \\  \hline
		9   & 261,080   & 2,111,013          &       13,026,161,682,466,252    & $\approx$ 9 $\times$ \#Subgraphs    \\   \hline
	\end{tabular}
\end{table*}

Here we are mainly interested in algorithms that count induced subgraphs, but non-induced subgraphs counting algorithms are also considered. Counting one or the other is equivalent since it is possible to obtain induced occurrences from non-induced occurrences, and vice-versa. However, we should note that, at the end of the counting process, induced occurrences need to be obtained by the algorithm. This choice penalizes non-induced subgraph counting algorithms since the transformation is quadratic on the number of subgraphs~\cite{floderus2015induced}. Some algorithms count orbits instead of subgraphs~\cite{hovcevar2014combinatorial}. However, counting orbits can be reduced to counting subgraphs and, therefore, these algorithms are also considered. 

We should note that we only consider the most common and well studied subgraph frequency definition, in which different occurrences might share a partial subset of nodes and edges, but there are other possible frequency concepts, in which this overlap is explicitly disallowed~\cite{schreiber2005frequency,elhesha2016identification}.

\subsection{Algorithms Not Considered}

In this work we focus on practical algorithms that are capable of counting all subgraphs of a given size. Therefore, algorithms that only target specific subgraphs are not considered (e.g., triads~\cite{schank2005finding}, cliques~\cite{finocchi2015clique, aliakbarpour2018sublinear}, stars~\cite{gonen2011counting,danisch2018listing} or subtrees~\cite{li2018mtmo}). Furthermore, given our focus on generalizability, we do not consider algorithms that are only capable of counting sugraphs in specific graphs (e.g., bipartite networks~\cite{sanei2018butterfly}, trees~\cite{czabarka2018number}), or that only count local subgraphs~\cite{dave2017clog}.

Graphs used throughout this work are simple, have a single layer of connectivity and do not distinguish the node or edge types with qualitative or quantitative features. Therefore we do not discuss here algorithms that use colored nodes or edges~\cite{guillemot2013finding,gholami2013rangi,ribeiro2014discovering}, and neither those that consider networks that are heterogeneous~\cite{gu2018homogeneous,rossi2019heterogeneous}, multilayer ~\cite{ren2019finding,boekhout2019efficiently}, labelled/attributed~\cite{mongiovi2018glabtrie}, probabilistic~\cite{sarkar2018new} or any kind of weighted graphs~\cite{williams2013finding}.

Finally, the networks we consider are static and do not change their topology. We should however note that there has been an increasing interest in temporal networks, that evolve over time~\cite{holme2012temporal}. Some algorithms beyond the scope of this survey try to tackle temporal subgraph counting, either by considering temporal networks as a series of static snapshots~\cite{hulovatyy2015exploring,aparicio2018graphlet}, by timestamping edges~\cite{paranjape2017motifs}, or by considering a stream of small updates to the graph topology~\cite{schiller2015stream,silva2017network,cannoodt2018incgraph,kallaugher2018sketching}.

\subsection{Applications and Related Problems}\label{sec:applications}

\subsubsection{Subgraph Isomorphism}
Given two graphs $G$ and $H$, the \textit{subgraph isomorphism} problem is the computational task of determining if $G$ contains a subgraph isomorphic to $H$. Although efficient solutions might be found for specific graph types (e.g., linear solutions exist for planar graphs~\cite{eppstein2002subgraph}), this is a known NP-Complete problem for general graphs~\cite{cook1971complexity}, and can be seen as much simpler version of counting, that is, determining if the number of occurrences is bigger than zero. This task is closely related to the \emph{graph isomorphism} problem~\cite{mckay1981practical, mckay2014practical}, that is, the task of determining if two given graphs are isomorphic. Since many subgraph counting approaches rely on finding the subgraphs contained in a large graph and then checking to what isomorphic class the subgraphs found belong to, subgraph isomorphism can be seen as an integral part of them. The well known and very fast \textsc{nauty} tool~\cite{mckay2003nauty} is used by several subgraph counting algorithms to assess the type of the subgraph found~\cite{wernicke2006fanmod,paredes2013towards,ribeiro2014g}.

\subsubsection{Subgraph Frequencies}

The small patterns found in large graphs can offer insights about the networks. By considering the frequency of all $k$-subgraphs, we have a very powerful and rich feature vector that characterizes the network. There has been a long tradition on using the triad census on the analysis of social networks~\cite{wasserman1994social}, and they have been used as early as in the 70s to describe local structure~\cite{holland1976local}. Examples of applications in this field include studying social capital features such as brokerage and closure~\cite{prell2008looking}, discovering social roles~\cite{doran2014triad}, seeing the effect of individual psychological differences on network structure~\cite{kalish2006psychological} or characterizing communication~\cite{uddin2013dyad} and social networks~\cite{charbey2019stars}. Given the ubiquity of graphs, these frequencies have also been used on many other domains, such as in biological~\cite{sole2007spontaneous}, transportation~\cite{wandelt2015evolution} or interfirm networks~\cite{madhavan2004two}.

\subsubsection{Network Motifs}

A subgraph is considered a \textit{network motif} if it is somehow exceptional. Instead of simply using a frequency vector, motif based approaches construct a \textit{significance profile} that associates an importance to each subgraph, typically related to how overrepresented it is. This concept first appeared in 2002 and it was first defined as subgraphs that occurred more often than expected when compared against a null model~\cite{milo2002network}. The most common null model is to keep the degree sequence and with this we can obtain characteristic network fingerprints that have been shown to be very rich and capable of classifying networks into distinct superfamilies~\cite{milo2004superfamilies}. Network motif analysis has since been in a vast range of applications, such as in the analysis of biological networks (e.g., brain~\cite{sporns2004motifs}, regulation and protein interaction~\cite{yeger2004network} or food webs~\cite{bascompte2005simple}), social networks (e.g., co-authorship~\cite{choobdar2012comparison} or online social networks~\cite{duma2014network}), sports analytics (e.g., football passing~\cite{bekkers2017flow}) or software networks (e.g., software architecture~\cite{valverde2005network} or function-call graphs~\cite{wu2018software}). 

In order to compute the significance profile of motifs in a graph $G$, most conceptual approaches rely on generating a large set of $R(G)$ of similar randomized networks that serve as the desired null model. Thus, subgraph counting needs to be performed both on the original network and on the set of randomized networks.  If the frequency of a subgraph $S$ is \emph{significantly bigger} in $G$ than it its average frequency in $R(G)$, we can consider $S$ to be a network motif of $G$~\cite{kashtan2004efficient}. Other approaches try to avoid exhaustive generation of random networks and, thus, avoid also counting subgraphs on them, by following a more analytical approach capable of providing estimations of the expected frequencies (e.g., using an expected degree model~\cite{picard2008assessing, schbath2008assessing,micale2018fast} or a scale-free model~\cite{stegehuis2019variational}. Nevertheless, there is always the need of counting subgraphs in the original network.

While network motifs are usually about induced subgraph occurrences~\cite{milo2002network,wong2012biological}, there are some motif algorithms that count non-induced occurrences instead~\cite{omidi2009moda,li2012netmode}. Moreover, although most of the network motifs usages assume the previously mentioned statistical view on significance as overrepresentation, there are other possible approaches~\cite{xia2019survey} such as using information theory concepts (e.g., motifs based on entropy~\cite{adami2011information,choodbdar2012weighted}, subgraph covers~\cite{wegner2014subgraph}, or minimum description length~\cite{bloem2017large}). We should also note that some approaches try to better navigate the space of "interesting" subgraphs, so that reaching larger motif sizes can be reached not by searching all possible larger $k$-subgraphs, but instead by leveraging computations of smaller motifs~\cite{patra2018motif,luo2018efficient}.

Finally, we should note that several authors use the term motif to refer to small subgraphs, even when it does not imply any significance value beyond simple frequency on the original network.

\subsubsection{Orbit-Aware Approaches and Network Alignment}

When authors use the term \textit{graphlet}, they commonly take orbits into consideration, and use metrics such as the graphlet-degree distribution (GDD, see details in section~\ref{sec:terminology}), a concept that appeared in 2007~\cite{przulj2007}. In this way, graphlet algorithms count how many times each node appears in each orbit. Unlike motifs, graphlets do not usually need a null model (i.e., networks are directly compared by comparing their respective GDDs). These orbit-aware distributions can be used for comparing networks. For instance, they have shown that protein interaction networks are more akin to random geometric graphs than to traditional scale-free networks~\cite{przulj2007}. Moreover, they are also used to compare nodes (using graphlet-degree vectors). This makes them useful for \textit{network alignment} tasks, where one needs to establish topological similarity between nodes from different networks~\cite{milenkovic2010optimal}. Several graphlet-based network alignment algorithms have been proposed and shown to work very well for aligning biological networks~\cite{kuchaiev2010topological,kuchaiev2011integrative,malod2015graal,sun2015simultaneous,aparicio2019temporal}.

\subsubsection{Frequent Subgraph Mining (FSM)}\label{sec:fsm}

FSM algorithms find subgraphs that have a \emph{support} higher than a given threshold. The most prevalent branch of FSM takes as input a bundle of networks and finds which subgraphs appear in a vast number of them - refereed to as \emph{graph transaction based FSM}~\cite{jiang2013survey}. These algorithms~\cite{yan2002gspan,huan2003efficient,nijssen2005gaston} heavily rely on the Downward Closure Property (DCP) to efficiently prune the search space. Algorithms for subgraph counting, which is our focus, can not, in general, rely on the DCP since it is not possible to know if growing an infrequent $k$-node subgraph will result, or not, in a frequent $k+1$ subgraph. Furthermore, we are not only interested in frequent subgraphs but in all of them, since rare subgraphs can also give information about the network's topology. A less prominent branch of FSM, \emph{single graph based FSM}, targets frequent subgraphs in a single large network, much like our subgraph counting problem. However, they adopt various support metrics that allow for the DCP to be verified, which, as stated previously, is not the case in the general subgraph counting problem~\cite{jiang2013survey}.
\

\subsection{Other Surveys and Related Work}

To the best of our knowledge there is no other comparable work to this survey in terms of scope, thoroughness and recency. Most of the already existing surveys that deal with subgraph counting are directly related to network motif discovery. Some of them are from before 2015 and therefore predate many of the most recent algorithmic advances~\cite{ribeiro2009strategies,wong2012biological,masoudi2012building,kim2013network,tran2014current}, and all of them only present a small subset of the strategies discussed here. There are more recent review papers, but they all differ from our work and have a much smaller scope. \citeAutRef{al2018triangle} only consider triangle counting, \citeAutRef{xia2019survey} focus mainly on significance metrics, and finally, while we here present a structured overview of more than 50 exact, approximate and parallel algorithmic approaches, ~\citeAutRef{jain2019network} presents a much simpler description of 5 different algorithms.

\section{Exact Counting}
\label{sec:exact}

As subgraph counting evolved over the years,
a multitude of algorithms and methods were developed that address the
problem in different ways and for distinct purposes. As such, it is
useful, although not easy, to group strategies together in order
to facilitate their understanding as well as learn why and how they
came about. With this in mind we divided this section into two major
groups of algorithms, namely enumeration and analytic approaches, which are further subdivided
in their respective section. Table~\ref{tab:over_exact} summarizes our proposed taxonomy composed 
of six aspects, ordered by their publication year: (i) \textbf{approach} (enumeration or analytic), 
(ii) \textbf{type} (a subgroup of the underlying approach), (iii) \textbf{$k$-restriction} (does the method only work
for certain subgraph sizes?), (iv) \textbf{orbit awareness} (does the method
also count orbits?), (v) \textbf{directed} (is the method applicable to directed graphs?) and (vi) if code is \textbf{publicly available}. At the end of this section,
we also present some related theoretical results that influenced some of the algorithms
we discuss.

\begin{table}[H]
	\small
	\centering
	\def\arraystretch{1.0}
	
	\caption{Overview of all major exact algorithms.}
	\label{tab:over_exact}
	\vspace{-0.2cm}
	\begin{tabular}{$l^c^c^c^c^c^c^c}
		\rowstyle{\bfseries}
		& Year & Approach & Type & $k$-restriction & Orbit & Directed & Code \\ \hline
		\textsc{Mfinder} \cite{milo2002network} & 2002 & Enum. & Classical & None & \xmark & \cmark & \cite{mfinder}\\
		\textsc{ESU} \cite{wernicke2005faster,wernicke2006fanmod} & 2005 & Enum. & Classical & None & \xmark & \cmark & \cite{fanmod} \\
		\textsc{Itzhack} \cite{itzhack2007optimal} & 2007 & Enum. & Classical & $\leq 5$ & \xmark & \cmark & \xmark \\
		\textsc{Grochow} \cite{grochow2007network} & 2007 & Enum. & Single-subgraph & None & \xmark & \cmark & \xmark \\
		\textsc{Kavosh} \cite{kashani2009kavosh} & 2009  & Enum. & Classical & None & \xmark & \cmark & \cite{kavosh}\\	
		\textsc{Gtries} \cite{ribeiro2010g,ribeiro2014g} & 2010 & Enum. & Encapsulation & None & \cmark & \cmark & \cite{gtries}\\	
		\textsc{Rage} \cite{marcus2010efficient,marcus2012rage} & 2010 & Analytic & Decomposition & $\leq 5$ & \xmark & \cmark & \cite{rage} \\
		\textsc{NeMo} \cite{koskas2011nemo} & 2011 & Enum. & Single-subgraph& None & \xmark & \cmark & \cite{nemo} \\
		\textsc{Netmode} \cite{li2012netmode} & 2012 & Enum. & Encapsulation & $\leq 6$ & \xmark & \cmark & \cite{netmode}\\
		\textsc{SCMD} \cite{wang2012symmetry} & 2012 & Enum. & Encapsulation & None & \xmark & \xmark & \xmark\\
		\textsc{acc-Motif} \cite{meira2012accelerated,meira2014acc} & 2012 & Analytic & Decomposition & $\leq 6$ & \xmark & \cmark & \cite{accmotif}\\
		\textsc{ISMAGS} \cite{demeyer2013index,houbraken2014index} & 2013 & Enum. & Single-subgraph & None & \xmark & \cmark & \cite{ismags}\\
		\textsc{Quatexelero} \cite{khakabimamaghani2013quatexelero} & 2013 & Enum. & Encapsulation & None & \xmark & \cmark & \cite{quatexelero}\\
		\textsc{FaSE} \cite{paredes2013towards} & 2013 & Enum. & Encapsulation & None & \xmark & \cmark & \cite{gtscanner} \\
		\textsc{ENSA} \cite{zhang2014motif} & 2014 & Enum. & Encapsulation & None & \xmark & \cmark & \xmark \\
		\textsc{Orca} \cite{hovcevar2014combinatorial,hovcevar2017combinatorial} & 2014 & Analytic & Matrix-based & $\leq 5$ & \cmark & \xmark & \cite{orca}\\
		\textsc{Hash-ESU} \cite{zhao2015hashesu} & 2015 & Enum. & Encapsulation & None & \xmark & \cmark & \xmark \\
		\textsc{Song} \cite{song2015method} & 2015 & Enum. & Encapsulation & None & \xmark & \cmark & \xmark \\
		\textsc{Ortmann} \cite{ortmann2016quad,ortmann2017efficient} & 2016 & Analytic & Matrix-based & $\leq 4$ & \cmark & \cmark & \xmark\\
		\textsc{PGD} \cite{ahmed2015efficient,ahmed2016estimation} & 2016 & Analytic & Decomposition & $\leq 4$ & \cmark & \xmark & \cite{PGD}\\
		\textsc{Patcomp} \cite{jain2017impact} & 2017 & Enum. & Encapsulation & None &  \xmark & \cmark & \xmark \\
		\textsc{Escape} \cite{pinar2017escape} & 2017 & Analytic & Decomposition & $\leq 5$ & \cmark & \xmark & \cite{escape} \\
		\textsc{Jesse} \cite{melckenbeeck2017,melckenbeeck2019optimising} & 2017 & Analytic & Matrix-based & None & \cmark & \xmark & \cite{jesse} \\
		\end{tabular}
		\end{table}

\subsection{Enumeration approaches}

A significant part of the history of practical subgraph counting algorithms is intertwined 
with network motif analysis. This is because when motifs were first
proposed~\cite{milo2002network}, they raised the interest and necessity for efficient subgraph counting, which has since been growing and establishing itself as a very important graph analysis primitive with multidisciplinary applicability.

Exact subgraph counting consists of \emph{counting} and \emph{categorizing} (i.e., determining the isomorphic class of) all subgraph occurrences.
Early methods \emph{first enumerate} all connected subgraphs with $k$-vertices and \emph{only afterwards
categorize} each subgraph found using a graph isomorphism tool like \textsc{nauty} \cite{mckay2003nauty}. We refer to these as {\bf  classical methods}.

Many methods followed this strategy, until new methods appeared that
counted the frequency of a single-subgraph category instead, thus avoiding the categorization step
necessary by the classical methods. This was done by only
enumerating one particular subgraph of interest. Even
though they were not the fastest methods for a network-centric
application, they were an important milestone towards the methods
that followed. We refer to these as {\bf single-subgraph-search methods}.

The next step was to combine the two previous ideas into a more efficient approach: 
merge the enumeration and categorization steps together. This
was achieved in different ways, such as using common topological
features of subgraphs or pre-computing some information about
subgraphs to avoid repeated computations of isomorphism. We refer to these as {\bf encapsulation methods}.

The next sections thoroughly delve into the most well-known methods of each
category, giving an historical perspective on each, in an effort to
understand each method's breakthroughs and drawbacks, and how subsequent algorithms built
upon them to reduce (or mitigate) their limitations. 
\subsubsection{Classical methods}
\label{sec:exact_class}

In the seminal work, \citeAutRef{milo2002network} first
defined the concept of network motif and also proposed \textsc{MFinder}, an
algorithm to count subgraphs. \textsc{MFinder}
is a recursive backtracking algorithm, that is applied to each
edge of the network. A given edge is initially stored on a set $S$, which is recursively
grown using edges that are not in $S$ but share one endpoint with at
least one edge in $S$. When $|S| = k$, the algorithm checks if
the subgraph induced by $S$ has been found for the first time by keeping a hash table of
subgraphs already found. If the subgraph was reached for the first time,
the algorithm categorizes it and updates the hash table (otherwise, the subgraph is ignored).

Another very important work, by \citeAutRef{wernicke2005faster}, proposed a new algorithm called \textsc{ESU}, also known
as \textsc{FANMOD} due to the graphical tool that uses \textsc{ESU} as its core algorithm~\cite{wernicke2006fanmod}. This
algorithm greatly improved on \textsc{MFinder} by never counting the same
subgraph twice, thus avoiding the need to store all subgraphs in a
hash table. \textsc{ESU} applies the same recursive method to each vertex $v$ of the
input graph $G$: it uses two sets $V_S$ and $V_E$, which initially are
set as $V_S = \{v\}$ and $V_E = N(v)$. Then, for each vertex $u$ in
$V_E$, it removes it from $V_E$ and makes $V_S = V_S \cup \{u\}$,
effectively adding it to the subgraph being enumerated and $V_E = V_E
\cup \{u \in N_{exc}(u, V_S) : L(u) > L(v)\}$ (where $v$ is the
original vertex to be added to $V_S$). The $N_{exc}$ here makes sure we only grow the list of
possibilities with vertices not already in $V_S$ and the condition
$L(u) > L(v)$ is used to break symmetries, consequently preventing any
subgraph from being found twice. This process is done several times
until $V_S$ has $k$ elements, which means $V_S$ contains a single
occurrence of a $k$-subgraph. At the end of the process, \textsc{ESU}
performs isomorphism tests to assess the category of each subgraph occurrence,
which is a considerable bottleneck.

\citeAutRef{itzhack2007optimal} proposed a new
algorithm that was able to count subgraphs using constant memory (in
relation to the size of the input graph). Itzhack et al. did not name their algorithm, 
so we will refer to it as \textsc{Itzhack} from here on.
\textsc{Itzhack} avoids explicitly computing the
isomorphism class of each counted subgraph by caching it for each
different adjacency matrix, seen as a bitstring. This strategy only works for
subgraphs of $k$ up to 5, since it would use too much memory for
higher values. Additionally, the enumeration algorithm is also different from \textsc{ESU}.
This method is based on counting all subgraphs that include a certain vertex, then removing
that node from the network and repeating the same procedure for the remaining nodes. For each vertex $v$, first
the algorithm considers the tree composed of the $k$ neighborhood of
$v$, that is, a tree of all vertices at a distance of $k - 1$ or less
from $v$. This is very similar to the tree obtained from performing a
breadth-first search starting on $v$, with the difference that vertices
that appear on previous levels of the tree are excluded if visited
again. This tree can be traversed in a way that avoids actually
creating it by following neighbors, and thus only using constant
memory. To perform the actual search, the method uses the concept of
{\it counting patterns}, which are different combinatorial ways of
choosing vertices from different levels of the tree. For instance, if
we are searching for 3-subgraphs, and considering that at the tree
root level we can only have one vertex, we could have the combinations
with pattern 1-2 (one vertex at root level 0, two vertices at level 1)
or with pattern 1-1-1 (one vertex at root level 0, one at level 1 and
one at level 2). In an analogous way, 4-subgraphs would lead to
patterns 1-1-1-1, 1-1-2, 1-2-1 and 1-3. Itzhack et al. claimed that \textsc{Itzhack} is over 1,000 times faster than \textsc{ESU}, however the
author of \textsc{ESU} disputed this claim in \cite{wernicke2011comment},
stating that the experimental setup was faulty and claimed that \textsc{Itzhack} is only
slightly faster than \textsc{ESU} (its speedup could  be attributed mainly to
the caching procedure). 

\citeAutRef{kashani2009kavosh} proposed a new algorithm
called \textsc{Kavosh}. Like \textsc{ESU} and \textsc{Itzhack}, the core idea of the \textsc{Kavosh} is to find all subgraphs that
include a particular vertex, then remove that vertex and continue from
there iteratively. Its functioning is very similar to that of \textsc{Itzhack}: it builds an implicit breadth-first search tree and then
uses a similar concept to the counting patterns used by \textsc{Itzhack}. However, it is a more general method since it does not perform
any caching of isomorphism information, allowing the enumeration of larger subgraphs.

\subsubsection{Single-subgraph-search methods}

The idea that it is possible to obtain a very efficient method of
counting a single subgraph category was first noticed by \citeAutRef{grochow2007network}. Their base
method consists on a backtracking algorithm that is applied to each
vertex. It tries to build a partial mapping from the input graph to
the target subgraph (the subgraph it is trying to count) by building
all possible assignments based on the number of neighbours. Grochow and Kellis also
suggested an improvement based on symmetry breaking, using
the automorphisms of the target subgraph to build set of conditions, of the form $L(a) < L(b)$, to
prevent the same subgraph from being counted multiple times. This symmetry breaking
idea allowed for considerable improvements in runtime, specially for
higher values of $k$. Grochow and Kellis did not name their algorithm, so we will refer to it as the \textsc{Grochow}
algorithm from here on.

\citeAutRef{koskas2011nemo} presented a new algorithm
which they called \textsc{NeMo}. This method draws some ideas from \textsc{Grochow}, 
since it performs a backtrack based search with symmetry
breaking in a similar fashion. Although, instead of using conditions
on vertex labels, it finds the orbits of the target subgraph and
forces an ordering between the labels of the vertices from the input
graph that match vertices in the target subgraph with the same
orbit. Additionally, it uses a few heuristics to prune the search
early, such as ordering the vertices from the target graph such that
for all $1 \leq i \leq k$, its first $i$ vertices are connected.

\textsc{ISMAGS}, which is based on its predecessor \textsc{ISMA}
\cite{demeyer2013index}, was proposed by \citeAutRef{houbraken2014index}. The base idea of this method is similar to
the one in \textsc{Grochow}, however, the authors use a clever node
ordering and other heuristics to speedup the partial mapping procedure.
Additionally, their symmetry breaking conditions are significantly improved by applying several heuristic techniques based on group
theory.

\subsubsection{Encapsulation methods}
\label{sec:exact_encap}

The ideas applied to \textsc{Grochow} introduced a way of escaping the
classic setup of enumerating and then categorizing subgraphs, albeit
focusing on a single subgraph. The next step would be to extend this
idea to a more general algorithm, which is appropriate to a full
subgraph counting. This was first done by \citeAutRef{ribeiro2010g} using a new data-structure they called the {\it
  g-trie}, for {\it graph trie}. The {\it g-trie} is a prefix tree for
graphs, each node represents a different graph, where the graph of a
parent node has shared common substructures with the graph of its
child node, which are characterized precisely by the vertices of the
graph of the child node. The root represents the one vertex graph with
one child, a node representing the edge graph, which in turn has two
children representing the triangle graph and the 3-path, and so
on. This tree can be augmented by giving each node symmetry breaking
conditions similar to those from \textsc{Grochow}. The authors show how to efficiently build this data-structure and
augment with the symmetry breaking conditions for any set of
graphs. Also, they describe a subgraph counting algorithm based on
using this data-structure along with an enumeration technique similar
to that of \textsc{Grochow}. However, since this data-structure
encapsulates the information of multiple graphs in an hierarchical
order, it achieves a much faster full subgraph counting algorithm. The usage of this data-structure has been significantly extended since
its original publication, such as a version for colored networks
\cite{ribeiro2014discovering} or an orbit aware version
\cite{aparicio2016extending}. A more detailed discussion of the data-structure and the subgraph
counting algorithm is presented in \cite{ribeiro2014g}. Also, even
though the subgraph counting algorithm was not named, we will refer to
it as the \textsc{Gtrie} algorithm from here on.

\textsc{Gtrie} encapsulates common topological information of the
subgraphs being counted, but there are other approaches, such as 
\citeAutRef{li2012netmode}, who developed \textsc{Netmode}. It builds on
\textsc{Kavosh}, by using its enumeration algorithm, but instead of using
\textsc{nauty} to perform the categorization step, it makes use of a
cache to store isomorphism information and thus is able to perform it
in constant time. This is very similar to what \textsc{Itzhack} does,
however, Li et al. suggested an improvement that allows \textsc{Netmode}
to scale to $k = 6$ without using too much memory. This improvement is
based on the {\it reconstruction conjecture}~\cite{harary1974survey}, that states that two
graphs with 3 or more vertices are isomorphic if their deck (the set of isomorphism classes of all vertex-deleted subgraphs of a graph) is the same. This is known to be false for directed graphs with $k = 6$, but
there are very few counter-examples that can be directly stored such as
in the $k \leq 5$ case, thus \textsc{Netmode} applies the conjecture for
all the remaining cases by building their deck, hashing its value and
storing its count in a table.

\citeAutRef{wang2012symmetry} proposed a new method 
called \textsc{SCMD} that counts subgraphs in compressed networks. \textsc{SCMD} applies a symmetry
compression method that finds sets of vertices that are in an
homeomorphism to cliques or empty subgraphs, which have the additional
property that any other vertex that connects to a vertex in the set is
connected to all other vertices in the set. These sets of vertices
form a partition of the graph that is obtained using a method
published in~\cite{macarthur2008symmetry}, which is based on looking
at vertices in the same orbit.
This
is a versatile method that can use
algorithms like \textsc{ESU} or \textsc{Kavosh} to enumerate all subgraphs of
sizes from 1 to $k$ in the compressed network. Finally, \textsc{SCMD} ``decompresses'' the results
by looking at all the different enumerated subgraphs and calculating
all the combinations that can form a decompressed subgraph. For
example, for $k = 3$, if a compressed 2-subgraph is found containing
two vertices: one compressed vertex representing a clique of 5
uncompressed vertices and a compressed vertex representing a single
vertex from the uncompressed graph, it results in $\binom{5}{2} +
\binom{5}{3}$ triangles from the uncompressed graph, obtained by
taking two vertices from the clique vertex and one from the other
vertex, which are all connected and thus form a triangle,
$\binom{5}{2}$, plus taking three vertices from the clique vertex
$\binom{5}{3}$. The authors argue that most complex networks exhibit high symmetries
and thus are improved by the application of this technique. Even
though their work only includes undirected graphs, the authors affirm
it is easy to extend the same concepts to directed networks. {\bf Xu et al.}
described another algorithm that enumerates subgraphs on compressed networks, called \textsc{ENSA}~
\cite{xu2014new,zhang2014motif}. Their method is based on an heuristic graph
isomorphism algorithm, and they also discuss an optimization based on identifying
vertices with unique degrees.

Following the ideas first applied in \textsc{Gtrie}, \citeAutRef{khakabimamaghani2013quatexelero} 
proposed a new algorithm they called \textsc{Quatexelero}. \textsc{Quatexelero} is built upon any incremental enumeration
algorithm, like \textsc{ESU}, and it implements a data structure similar
to a quaternary tree. Each node in the tree represents a graph, that
can be built by looking into the nodes from the path from it to the
root of the tree. Additionally, all graphs represented by a single
node belong to the same isomorphism class. To fill the tree, initially a pointer to the root of the tree is
set. Whenever a new vertex is added to the partial enumeration map,
\textsc{Quatexelero} looks into the existing edges between the newly
added node and the previously existing nodes in the mapping and stores
its information in the quaternary tree. For each vertex in the
mapping, depending on whether there is no edge, an inedge, an outedge
or a biedge between it and the newly added vertex, the pointer is
assigned to one of its four children, creating it if it was
nonexistent.

Parallel to the publishing of the work of \textsc{Quatexelero}, \citeAutRef{paredes2013towards} proposed \textsc{FaSE}.
The idea of \textsc{FaSE} is similar to the
one from \textsc{Quatexelero}, however, instead of using a quaternary
tree, it uses a data-structure similar to the g-trie, albeit without
the symmetry breaking condition augmentation. This data-structure has
the same property as the quaternary tree that every node represents a
graph and each node is built using the adjacency information of a
newly added vertex in relation to the vertices present in its parent.

Other works that extend these ideas have been proposed
subsequently. For example, \citeAutRef{zhao2015hashesu}  propose \textsc{Hash-ESU}, an
algorithm based on the same idea from \textsc{Quatexelero} and \textsc{FaSE}, but which hashes the adjacency
information instead of storing it in a tree. Another example is the work by \citeAutRef{song2015method}. They describe a method that starts by
enumerating all $k = 3$ subgraphs using \textsc{ESU} and then use dynamic
programming to grow connected sets and perform the counting. Their
algorithm was not named, so we will refer to it as the \textsc{Song}
algorithm from here on.

Both \textsc{Quatexelero} and \textsc{FaSE} have potential memory issues,
since there may be several nodes representing the same graph, which is
not a problem for \textsc{Gtrie} since it only stores one copy of each
possible graph. To address this, \citeAutRef{jain2017impact} proposed \textsc{Patcomp}. Their method compresses the
quaternary tree using a technique similar to a radix tree, however,
their method is 2 to 3 times slower and only saves around $10\%$ of
the memory usage.

\subsection{Analytic approaches}\label{sec:anal}

Since the overall goal of the problem we are aiming to solve is to
count subgraphs, it is not necessary to explicitly enumerate each
connected set of size $k$. Here lies the difference between {\it
  counting} and {\it enumerating} or {\it listing}. It was with this
in mind that a new class of methods emerged striving to avoid enumerating
all subgraphs in a graph. We can point two main approaches to this
type of counting.

The first one tries to relate the frequency of each subgraph with the
frequencies of other subgraphs of the same or smaller size. This
permits constructing a matrix of linear equations between subgraphs
frequencies that can be solved using traditional linear algebra
methods. We refer to these as {\bf matrix based methods}.

The second approach targets each subgraph individually by decomposing it in
several smaller patterns of graph properties, like common neighbors,
or triangles that touch two vertices. We refer to these as {\bf
  decomposition methods}.

\subsubsection{Matrix based methods}

The first known method to apply a practical analytic approach based on
matrix multiplication to subgraph counting was \textsc{ORCA}, a work by
\citeAutRef{hovcevar2014combinatorial}, which is based on counting orbits
and not directly subgraphs. Their original work was targeted at orbits
in subgraphs up to 5 vertices and, because of that, they count
induced subgraphs specifically, while most analytic approaches count non-induced occurrences. 
\textsc{ORCA} works by setting up a system of linear equations per vertex of the input graph that relate different orbit frequencies, which are the system's variables. This system of linear equations contains information about the input graph. By construction, the matrix has a
rank equal to the number of orbits minus 1, thus to solve it one only
need to find the value of one the orbit frequencies and use any
standard linear algebra method to solve it. Usually, the orbit
pertaining to the clique is chosen, since there are efficient
algorithms to count this orbit and, for sparse enough networks, it is usually the one with the
least occurrences, making it less expensive to count.

Later, the authors of \textsc{ORCA} extended their work by suggesting a
way of producing equations for arbitrary sized subgraphs~\cite{hovcevar2017combinatorial}, although their available practical implementation is still limited to size 5~\cite{orca}. Another possible extension for \textsc{ORCA} was proposed by~\cite{melckenbeeck2017} with the \textsc{Jesse} algorithm, which was further complemented with a strategy for optimizing the computation by carefully selecting less expensive equations~\cite{melckenbeeck2019optimising}. 

Similar to \textsc{ORCA}, but using a different strategy, \citeAutRef{ortmann2016quad} proposed a new method, which
they further improved and better described in
\cite{ortmann2017efficient}. They also target orbits, but for
subgraphs of size up to 4. Their approach is based on looking into
non-induced subgraphs using them to build linear equations that are
less expensive to compute. Additionally, they also apply an improved
clique counting algorithm. \citeAutRef{ortmann2016quad} did not name their algorithm, so we will refer to it as the \textsc{Ortmann}
algorithm from here on.

\subsubsection{Decomposition methods}

Before \textsc{ORCA} was proposed, the first ever practical method that
used an analytic approach to subgraph counting was \textsc{Rage}, by {\bf
  Marcus and Shavitt}~\cite{marcus2010efficient,marcus2012rage}. Their method is based on~\cite{gonen2009approximating}
which employs similar techniques but with a more theoretical
focus. \textsc{Rage} targets non-induced subgraphs and orbits of size 3
and 4. It does so by running a different algorithm for each of the 8
existing subgraphs. Each algorithm is based on merging the
neighborhoods of pairs of vertices to ensure that a given quartet of
vertices have the desired edges to form a certain subgraph.

\textsc{acc-Motif}, which was proposed by \citeAutRef{meira2012accelerated}
and then further improved in
\cite{meira2014acc}, was also one of the first methods to employ an
analytic strategy, but stands out as the only known analytic method
that also works for directed subgraphs. \textsc{acc-Motif} also targets
non-induced subgraphs and their latest version supports up to size 6
subgraphs.

Another method that followed this trend of decomposition methods is
\textsc{PGD}, proposed by {\bf Ahmed et al.}~\cite{ahmed2015efficient,ahmed2017graphlet}. This method builds on
the classic triangle counting algorithm to count several primitives
that are then used to obtain the frequency of each subgraph and
orbit. It is currently one of the fastest methods, however it can only
count undirected subgraphs of size 3 and 4. Additionally, as most
analytic methods, it is highly parallelizible. Due to its versatile nature, \textsc{PGD} has been expanded to other
frequency metrics and it stands out as one of the only available
efficient methods that can count motifs incident to a vertex or edge
of the graph~\cite{ahmed2016estimation}, in what is called a ``local
subgraph count''.

More recently, \textsc{ESCAPE} was proposed by \citeAutRef{pinar2017escape}. This method is based on a divide and conquer
approach that identifies substructures of each counting subgraph to
partition them into smaller patterns. It is a very general method, but
with the correct choices for decomposition, it is possible to describe
a set of formulas to compute the frequency of each subgraph. The
original paper only describes the resulting formulas to subgraphs up
to size 5, however larger sizes can be obtained with some effort. As
of this writing, it is possibly the most efficient algorithm to count
undirected subgraphs and orbits up to size 5.

\subsection{Theoretical Results}

Even though the focus of this work is to look at the proposed
practical algorithms, it is
important to note that some of the existing work drew
inspiration from numerous more theoretical-oriented works. Thus, it is
of relevance to briefly summarize some of the achievements in this
area and we will do so with a special interest in those that directly
influenced some of the algorithms discussed in this section.

The first interest in subgraph counting stemmed from the world of
enumeration algorithms. The book ``Enumeration in
Graphs''~\cite{bezem1987enumeration} surveyed several methods to
enumerate several different structures in a graph, such as cycles,
trees or cliques. Even though these are specific subpatterns, they
often represent the fundamental computation that needs to be done
in order to enumerate any subgraph. These ideas were translated into
works that count
subgraphs by efficiently enumerating simpler substructures like
these~\cite{kloks2000finding,itzhack2007optimal}. Approximation schemes can also be developed with this in mind, which approximates the frequency
of several subgraph families like cycles or paths and then generalize
these for all size 4 subgraphs~\cite{gonen2009approximating}.

Another example of an initially purely theoretical technique is the work by~\citeAutRef{kowaluk2013counting}, which was one of the inspirations for the
multitude of matrix based analytic algorithms for counting
subgraphs. In fact, the most efficient algorithms are
based on several theoretical foundations that allow a tighter analysis of
runtime. Due to this interplay, it is worth mentioned a few more recent papers
on subgraph counting and enumerating. There is an interest in
finding efficient algorithms that are parameterized or sensitive to certain properties of
the graph, such as independent sets~\cite{williams2013finding} or its maximum degree~\cite{bjorklund2018counting}. Another current interest is in counting
and enumerating subgraphs in a dynamic or online environment~\cite{lin2012arboricity}. Finally, another active theoretical
topic is to find optimal algorithms for
enumeration, as in~\cite{ferreira2013efficiently}, as
well as proving lower bounds on their time complexity, as~\citeAutRef{bjorklund2014listing} does for triangle listing.

\section{Approximate Counting}\label{sec:sampling}

Despite the significant advances made towards faster subgraph counting algorithms, current state of the art
algorithms that determine exact frequencies still take hours, if not days, for very large networks.
With the ever increasing amount of data our society generates (e.g., in big social networks such as Twitter
and Facebook new members/nodes join every second), it is unfeasible to count all possible subgraphs. To solve
this problem, subgraph counting research drifted towards approximating these frequencies, making a trade-off between
losing accuracy but gaining time. Additionally, in some applications, approximate subgraphs counts might be sufficient~\cite{kashtan2004efficient,ribeiro2010estimation}.

Throughout this section we make a distinction between algorithms that estimate (i) subgraph frequencies or (ii) subgraph concentrations. Estimating subgraph frequencies is harder since the algorithm needs
to know the magnitude of the values, whereas to estimate concentrations the algorithm only needs to know the different
proportions of each subgraph in the network. Obtaining subgraph concentrations from subgraph frequencies is trivial but the reverse requires extra computational tasks.

We further split the approximate counting algorithms in five broad categories: \textbf{randomised enumeration}, \textbf{enumerate-generalize}, \textbf{path sampling},
\textbf{random walk}, and \textbf{colour coding}. In each subsection, we provide an algorithmic overview of each strategy and delve into
the individual algorithms that implement it and how they differ between themselves.

Tables~\ref{tab:approx_algs} and~\ref{tab:restaccess_algs} summarize the algorithms we discuss in the section. We split the methods into
algorithms where the
full topology is assumed (Table~\ref{tab:approx_algs}) and algorithms tailored to networks with restricted access (Table~\ref{tab:restaccess_algs}). Although some algorithms from each category may work in the other setting,
they excel for the task they were designed for and the distinction should be made clear. The tables summarize our proposed taxonomy composed 
of five aspects, ordered by their publication year: (i) the type of {\bf output} (frequencies or
concentrations), (ii) \textbf{$k$-restrictions} (does the method only work
for certain subgraph sizes?), (iii) \textbf{directed} (is the method applicable to directed graphs?), (iv) the {\bf strategy} it employs, according to
our taxonomy, and (v) if code is \textbf{publicly available}. Note that some authors do not have 
executable versions publicly available, but will be happy to share them through email. We mark these algorithms with a \cmark in the code
column of the table.

\begin{table}[!h]
        \small 
        \centering
        \def\arraystretch{1.0}
        \caption{Algorithms for approximate subgraph counting.}
        \label{tab:approx_algs}
        \begin{tabular}{$l^l^l^c^c^l^c }
                \rowstyle{\bfseries}
                & Year & Output & $k$-restriction & Directed & Strategy & Code\\
                & & & & \\[-8pt] 
                         \hline
                         \textsc{ESA}~\cite{kashtan2004efficient}  & 2004 & Conc. & None & \cmark & Random Walk & \cite{mfinder}  \\[2pt]
                         \textsc{RAND-ESU}~\cite{wernicke2005faster}  & 2005 & Freq. & None & \cmark & Rand. Enum. & \cite{fanmod}  \\[2pt]
                         \textsc{TNP}~\cite{prvzulj2006efficient}  & 2006 & Conc. & 5 & \xmark & Enum. - Generalize & \xmark  \\[2pt]
                         \textsc{RAND-GTrie}~\cite{ribeiro2010estimation}  & 2010 & Freq. & None & \cmark & Rand. Enum. & \cite{gtries}\\[2pt]
                         \textsc{GUISE}~\cite{bhuiyan2012guise}  & 2012 & Conc. & 5 & \xmark & Random Walk & \cite{guise}  \\[2pt]
                         \textsc{RAND-SCMD}~\cite{wang2012symmetry}  & 2012 & Freq. & None & \cmark & Enum. - Generalize & \xmark  \\[2pt]
                         \textsc{Wedge Sampling}~\cite{seshadhri2013triadic}  & 2013 & Freq. & 3 & \cmark & Path Sampling & \cite{wedgesamp}  \\[2pt]
                         \textsc{GRAFT}~\cite{rahman2014graft}  & 2014 & Freq. & 5 & \xmark & Enum. - Generalize & \cite{graft}  \\[2pt]
                         \textsc{PSRW \& MSS}~\cite{wang2014efficiently}  & 2014 & Conc. & None & \xmark & Random Walk & \xmark  \\[2pt]
                         \textsc{MHRW}~\cite{saha2015finding}  & 2015 & Conc. & None & \xmark & Random Walk & \cmark  \\[2pt]
                         \textsc{RAND-FaSE}~\cite{paredes2015rand}  & 2015 & Freq. & None & \cmark & Rand. Enum. & \cite{fase}  \\[2pt]
                         \textsc{Path Sampling}~\cite{jha2015path}  & 2015 & Freq. & 4 & \xmark & Path Sampling & \xmark  \\[2pt]
                         \textsc{$k$-profile sparsifier}~\cite{elenberg2015beyond,elenberg2016distributed}  & 2016 & Freq. & 4 & \xmark & Enum. - Generalize & \cite{eelenberggit} \\[2pt]
                         \textsc{MOSS}~\cite{wang2018moss} & 2018 & Freq. & 5 & \xmark & Path Sampling & \cite{moss}  \\[2pt]
                         \textsc{SSRW}~\cite{yang2018ssrw} & 2018 & Freq. & 7 & \xmark & Random Walk & \xmark \\[2pt]
                         \textsc{CC}~\cite{bressan2018motif} & 2018 & Freq. & None & \xmark & Color Coding & \cite{bressancc} \\[2pt]
        \end{tabular}
\end{table}

\begin{table}[!h]
        \small 
        \centering
        \def\arraystretch{1.0}
        \caption{Algorithms for approximate subgraph counting with restricted access.}
        \label{tab:restaccess_algs}
        \begin{tabular}{$l^l^l^c^c^l^c }
                \rowstyle{\bfseries}
                & Year & Output & $k$-restriction & Directed & Strategy & Code\\
                & & & &  \\[-8pt]
                \hline
                         \textsc{WRW}~\cite{han2016waddling}  & 2016 & Conc. & None & \xmark & Random Walk & \xmark  \\[2pt]
                         \textsc{IMPR}~\cite{chen2016mining}  & 2016 & Freq. & 5 & \xmark & Random Walk & \cite{impr}  \\[2pt]
                         \textsc{CSS} \& \textsc{NB-SRW}~\cite{chen2016general}  & 2016 & Conc. & None & \xmark & Random Walk & \cmark  \\[2pt]
                         \textsc{Minfer}~\cite{wang2017inferring}  & 2017 & Conc. & 5 & \cmark & Enumerate - Generalize & \xmark  \\[2pt]
        \end{tabular}
\end{table}

\subsection{Randomised Enumeration}

These algorithms are adaptations of older enumeration algorithms that perform exact counting. They have the particularity
that they all induce a tree-like search space in the computation, where the leaves are the subgraph occurrences, and thus perform the approximation
in a similar manner. Each level of the search
tree is assigned a value, $p_i$, which denotes the probability of transitioning from parent
node to the child node in the tree. In this scheme, each leaf in this tree is reachable with
probability $P = \prod_{i=1}^{k} p_i$ and the frequency of each subgraph is estimated using the number of samples
obtained of that subgraph divided by $P$.

\begin{figure}[h]
\includegraphics[width=0.8\textwidth]{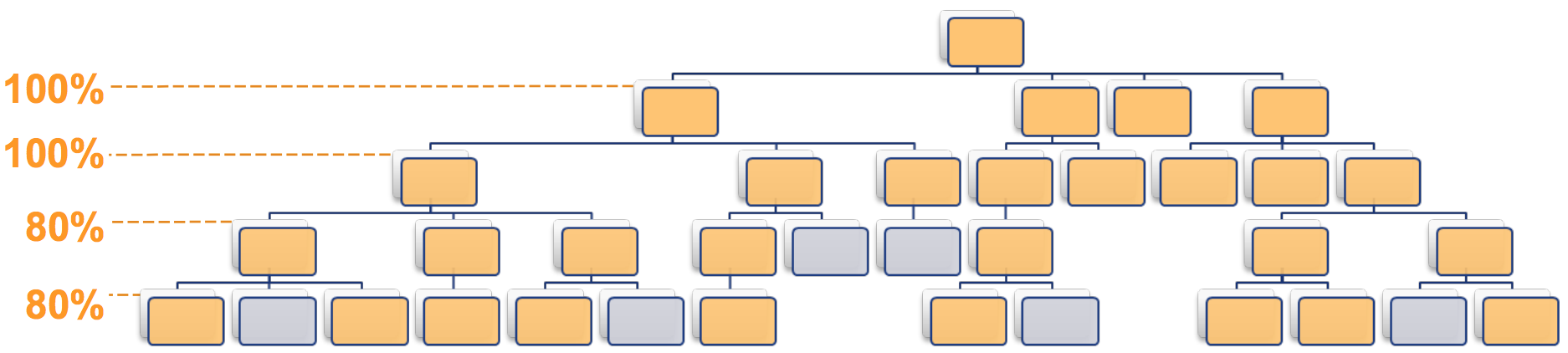}
\vspace{-0.35cm}
\caption{Example of a tree-like search space induced by a Randomised Enumeration algorithm and a possible distribution of transition probabilities per tree level.}
\label{fig:trees_probs_samp}
\end{figure}

Figure~\ref{fig:trees_probs_samp} illustrates how probabilities are added to the search tree. In this specific example,
which could be equivalent to searching subgraphs of size 4, the first two levels of the tree have probability 100\%, so their successors
are all explored. On the other hand, in the last two levels, the probability of exploring a node in the tree is only 80\%, therefore some
nodes, marked as grey, are not visited.

The first algorithm to implement this strategy was \textsc{RAND-ESU} by~\citeAutRef{wernicke2005faster}, an approximate
version of \textsc{ESU} (described in Section~\ref{sec:exact_class}). Recall that \textsc{ESU} maintains two sets $V_S$ and
$V_E$, the set of vertices in the subgraph and the set of candidate vertices for extending the subgraph. When adding a vertex 
from $V_E$ to $V_S$, this vertex is added with probability $p_{|V_S|}$, where $|V_S|$ is the depth of the search tree.

Using the more efficient {\it g-trie} data structure,~\citeAutRef{ribeiro2010estimation} proposed
\textsc{RAND-GTrie} and~\citeAutRef{paredes2015rand} proposed \textsc{RAND-FaSE}. Each level of the {\it g-trie} is assigned a probability, $p_i$.
When adding a new vertex to a subgraph of size $d$, corresponding to depth $d$ in the {\it g-trie}, this is done with
probability $p_d$.

\subsection{Enumerate-Generalize}

The general idea of these algorithms is to perform an exact count on a smaller network that was obtained from the original one (e.g., a sample, or a compressed network).
From the frequencies of each subgraph in the smaller network, the frequencies in the original network are estimated. Algorithms vary on (i) how the smaller network is obtained and on (ii) which estimator they use.

The first example of an algorithm in this category is Targeted Node Processing (\textsc{TNP}) by~\citeAutRef{prvzulj2006efficient}.
This algorithm is specially tailored for protein-protein interaction
,that, according to the authors,
have a periphery that is sparser than the more central parts of the network. Using this information, it performs an exact count
of the subgraphs in the periphery of the network and uses their frequencies to estimate the frequencies in the rest
of the network. The authors claim that, due to the uniformity of the aforementioned networks, the distribution of the subgraphs
in the fringe is representative of the distribution in the rest of the network.

\textsc{SCMD} by \citeAutRef{wang2012symmetry} (already covered in Section~\ref{sec:exact_encap})
allows the use of any approximate counting method in the compressed graph. There is no guarantee that subgraphs are
counted uniformly in the compressed graph, introducing a bias that needs to be corrected. The authors give the example of this bias
when using their method in conjunction with \textsc{RAND-ESU}. If each leaf (subgraph)
of depth $k$ in the search tree is reached with probability $P$ and a specific subgraph in the compressed graph is sampled
with probability $\rho$, then, to correct the sampling bias, the probability of decompressing the relevant $k$-subgraph is $P/\rho$.

In \textsc{GRAFT}, \citeAutRef{rahman2014graft} provide a strategy for counting undirected graphlets of size up to 5, using edge sampling.
The algorithm starts by picking an edge $e_g$ from each of the 29 graphlets and a set of edges sampled
from the graph $\mathcal{S}$, without replacement. For each edge $e \in \mathcal{S}$ and for each graphlet $g$, the frequency of $g$ is 
calculated such that $e$ has the same position in $g$ as $e_g$ ($e$ is said to be aligned with $e_g$). These frequencies are summed for all
edges and divided by a normalising factor, based on the automorphisms of each graphlet, which becomes the estimation for the
frequency of that graphlet in the whole network. Note that if $\mathcal{S}$ is equal to $E(G)$, the algorithm outputs an exact answer.

\citeAut{elenberg2015beyond} create estimators for the frequency of size 3~\cite{elenberg2015beyond} and 4~\cite{elenberg2016distributed}
subgraphs. A major difference from this work to previous ones is that \citeAut{elenberg2015beyond} estimate the frequencies of subgraphs
that are not connected, besides the usual connected ones. The authors start by removing each edge from the network with a certain probability 
and computing the exact counts in this ``sub-sampled'' network. Then, they craft a set of linear equations that relate the exact
counts on this smaller network to the ones of the original network. Using these equations, the estimation of the frequency
of the subgraphs in the original network follows.

\citeAutRef{wang2017inferring} introduce an algorithm that aims to estimate the subgraph concentrations of a network
when only a fraction of its edges are known. They call this a ``RESampled Graph'', obtained from the real network through
random edge sampling, a common scenario on applications such as network traffic analysis. A key aspect of this algorithm is the number of non-
induced subgraphs of a size $k$ graphlet that are isomorphic to another size $k$ graphlet, an example of this calculation can be found in
Table~\ref{tab:noninduc_rel4}. Using this number and the proportion of edges
sampled to form the smaller network,  the authors compute the probability that a subgraph in the ``RESampled Graph'' is isomorphic to
another subgraph in the original graph. Then, an exact counting algorithm is applied to the ``RESampled Graph'' and by composing the
results from this algorithm with the aforementioned probability, the subgraph concentrations in the original network are estimated.

\subsection{Path Sampling}

\begin{figure}[!ht]
  \includegraphics[width=0.85\textwidth]{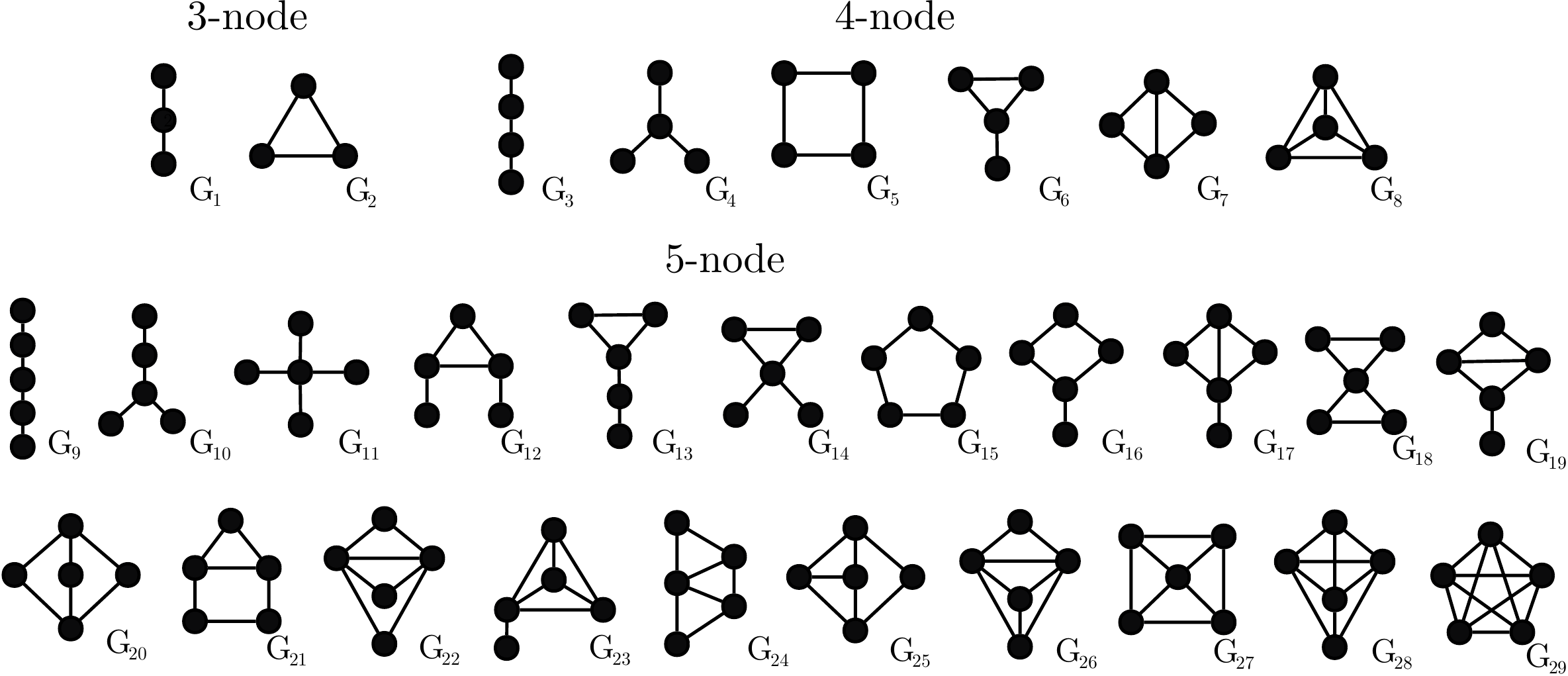}
  \vspace{-0.35cm}
  \caption{The 29 isomorphic classes of undirected subgraphs between size 3 and 5.}
  \label{fig:345undirgraphlets}
\end{figure}

This family of algorithms relies on the idea of sampling path subgraphs to estimate the frequencies of the other subgraphs. Path
subgraphs are composed by 2 exterior nodes and $k-2$ interior nodes (where $k$ is the size of the subgraph) arranged in a single
line; the interior nodes all have degree of 2, while the exterior nodes have degree of 1. Examples of these are the subgraphs $G_1$, $G_3$
and $G_9$ in Figure~\ref{fig:345undirgraphlets}. The main idea for these algorithms, mainly for $k \geq 4$, is relating the number of
non-induced occurrences of each subgraph of size $k$ in the other size $k$ subgraph. For example, when $k = 4$, there are 4 non-induced
occurrences of $G_3$ in $G_5$ or 12 non-induced occurrences of $G_3$ in $G_8$. Table~\ref{tab:noninduc_rel4} shows this full
relationship when $k = 4$.

\begin{table}
  \begin{center}
    \begin{tabular}{|c||c|c|c|c|c|c|}
      \hline & $g_3$ & $g_4 $ & $g_5$ & $g_6$ & $g_7$ & $g_8$\\
      \hline
      \hline $g_3$ & 0 & 1 & 2 & 4 & 6 & 12\\
      \hline $g_4$ & 1 & 0 & 1 & 0 & 2 & 4\\
      \hline $g_5$ & 0 & 0 & 0 & 1 & 1 & 3\\
      \hline $g_6$ & 0 & 0 & 1 & 0 & 4 & 12\\
      \hline $g_7$ & 0 & 0 & 0 & 0 & 1 & 6\\
      \hline $g_8$ & 0 & 0 & 0 & 0 & 0 & 1\\
      \hline
    \end{tabular}
    \caption{Number of non-induced occurences of each undirected graph of size 4 in each other. Position $(i,j)$ in the table indicates the number of times that graph $i$ occurs non-induced in graph $j$.}
    \label{tab:noninduc_rel4}
  \end{center}
\end{table}

\citeAutRef{seshadhri2013triadic} introduced the idea of \emph{wedge sampling}, where wedges denote size 3 path subgraphs. The
premise of the algorithm is simple, they select a number of wedges uniformly at random and check whether they are closed or not. 
The fraction of closed wedges sampled is an estimation from for the clustering coefficient, from which the number of triangles can be
derived.

Building on the idea of \emph{wedge sampling},~\citeAutRef{jha2015path} propose \emph{path sampling} to estimate
the frequency of size 4 graphlets. The main primitive of the algorithm is sampling non-induced
occurrences of $G_3$ and determining which graphlet is induced by that sample. The estimator relies on both the number of induced
subgraphs counted via the sampling and information contained in Table~\ref{tab:noninduc_rel4}. Finally, the authors
determine an equation to count the number of stars with 4 nodes ($G_4$) based on the frequencies of each other graphlet, since $G_4$
does not have any non-induced occurrence of $G_3$.

Applying the same concepts to size 5 subgraphs, \citeAutRef{wang2018moss} present \textsc{MOSS}-5. For size 5,
sampling paths is not enough to estimate the frequencies of all different subgraphs, as there are 3 subgraphs
that do not have a non-induced occurrence of a path: $G_{10}$, $G_{11}$ and $G_{14}$. On the other hand,
$G_{11}$ does not have a non-induced occurrence in 3 subgraphs as well ($G_9$, $G_{10}$ and $G_{15}$). Using this knowledge, the authors
create an algorithm divided in two parts: first it samples non-induced size 5 paths ($G_{9}$), similarly to~\citeAutRef{jha2015path},
and then they repeat the procedure but sampling occurrences of $G_{11}$ instead. Combining the results from these two sampling schemes,
the authors are able to estimate the frequency of every size 5 subgraph.

To the best of our knowledge, \textsc{MOSS}-5 is the
algorithm that achieves the best trade-off of accuracy and time to estimate the frequency of $5$-subgraphs, as it is able
to reach very small errors (magnitude $10^{-2}$) with a very limited number of samples, even for big networks. However
the ideas behind \textsc{MOSS}-5 are not easily extendable to directed subgraphs and larger sized undirected subgraphs due
to the ever increasing number of dependencies between the number of non-induced occurrences, making it harder to use the information
contained in a table similar to Table~\ref{tab:noninduc_rel4} for these cases.

\subsection{Random Walk}

A random walk in a graph $G$ is a sequence of nodes, $R$, of the form $R = (n_1, n_2, \ldots)$, where $n_1$ is the seed node and
$n_i$ the $i$th node visited in the walk. A random walk can also be seen as a Markov chain. We identify two main approaches
to sample subgraphs using random walks. The first is incrementing the size of the walk until a sequence of $k$ distinct nodes is drawn,
forming a $k$-subgraph, which is then identified by an isomorphism test. The second approach is considering a graph of relationships
between subgraphs, where two subgraphs are connected if one can be obtained from the other by adding or removing a node or
by adding or removing an edge. A random walk is then performed on this graph instead of on the original one.

\citeAutRef{kashtan2004efficient}, in their seminal work commonly called \textsc{ESA} (Edge Sampling), implemented
one of the first subgraph sampling methods in the \textsc{MFinder} software. The authors propose to
do a random walk on the graph, sampling one edge at a time until a set of $k$ nodes is found, from which
the subgraph induced by that set of nodes is discovered. This method resulted in a biased estimator. To
correct the bias, the authors propose to re-weight the sample, which takes exponential time in the
size of the subgraphs.

\citeAutRef{bhuiyan2012guise} develop \textsc{GUISE} that computes the graphlet degree distribution
for subgraphs of size 3, 4 and 5 in undirected networks. The algorithm is based on Monte Carlo Markov Chain (MCMC) sampling.
It works by sampling a seed graphlet, calculating its neighbourhood (a set of other graphlets), picking
one randomly and calculating an acceptance probability to transition to this new graphlet. This process is then
repeated until a predefined number of samples is taken from the graph. The neighbourhood of a graphlet
is similar to the graph of relationships previously mentioned, but to obtain a $k$-graphlet from another $k$-graphlet, a node
from the original one is removed and, if the remaining $k-1$ nodes are connected, their adjacency lists are concatenated
and nodes are picked from there to form the new $k$-graphlet. 

A similar approach to \textsc{GUISE} is used 
by~\citeAutRef{saha2015finding}, where MCMC sampling is also used to compute subgraph concentration. A difference
to \textsc{GUISE} is that the size of graphlets is theoretically unbound and only a specific size $k$ is counted,
whereas \textsc{GUISE} counts graphlets of size 3, 4 and 5 simultaneously. They also suggest a modified version where
the acceptance probability is always one (that is, there is always a transition to the new subgraph), which introduces a bias
towards graphlets with nodes with a high degree. In turn, they propose an estimator that re-weights the concentration
to remove this bias.

\citeAutRef{wang2014efficiently} propose a random walk based method to estimate subgraph concentrations that aims
to improve on the approach taken by \textsc{GUISE}. The main improvement over \textsc{GUISE} is that no samples
are rejected, avoiding a cost of sampling without any gain of information. The authors use a graph of relationships between
connected induced subgraphs, where two $k$-subgraphs are connected if they share $k-1$ nodes, but this graph is not
explicitly built, reducing memory costs. The basic algorithm is just a simple random walk over this graph of relationships.
The authors also present two improvements: \emph{Pairwise Subgraph Random Walk} (\textsc{PSRW}), estimates
size $k$ subgraph by looking at the graph of relationships composed by $k-1$-subgraphs; \emph{Mixed Subgraph Sampling} (\textsc{MSS}),
estimates subgraphs of size $k-1$, $k$ and $k+1$ simultaneously.

\citeAutRef{han2016waddling} present an algorithm to estimate subgraph concentration based on random walks. Their
algorithm, \emph{Waddling Random Walk} (\textsc{WRW}), gets its name from how the random walk is performed, allowing sampling of
nodes not only on the path of the walk, but also query random nodes in the neighbourhood. Let $l$ be the number of
vertices (with repetition) in the shortest path of a particular $k$-graphlet. The goal of the waddling is to reduce
the number of steps the walk has to take to identify graphlets with $l > k$. While executing a random walk to identify
a $k$-subgraph, the waddling approach limits the number of nodes explored to the size of the subgraph, $k$.

\citeAutRef{chen2016mining} propose a random walk based algorithm to estimate graphlet counts in online social networks,
which are often restricted and the entire topology is hidden behind a prohibitive query cost.
With this context in mind, the authors introduced the concepts of \emph{touched} and \emph{visible} subgraphs. The former
are subgraphs composed of vertices whose neighbourhood is accessible. The latter possess one and only one vertex with
inaccessible neighbourhood. Their method, \textsc{IMPR}, works by generating $k-1$-node \emph{touched} subgraphs
via random walk and combining them with their node's neighbourhood for obtain $k$-node \emph{visible} subgraphs, 
which form the $k$-node samples.

\citeAutRef{chen2016general} introduce a new framework that incorporates \textsc{PSRW} as a special case.
To sample $k$-subgraphs, the authors also use a graph of relationships between
connected induced $d$-subgraphs, $d \in \{1,..,k-1\}$, and perform a random walk over this graph. The difference
to \emph{PSRW} is that \emph{PSRW} only uses $d$ as $k-1$, which becomes ineffective as $k$ grows to larger sizes.
The authors also augment this method of sampling with a different re-weight coefficient to improve estimation accuracy
and add non-backtracking random walks, which eliminates invalid states in the Markov Chain that do not contribute to the estimation.

\citeAutRef{yang2018ssrw} introduce another algorithm using random walks, \emph{Scalable subgraph Sampling via Random Walk}
(\textsc{SSRW}), able to compute both frequencies and concentrations of undirected subgraphs of size up to 7. The next nodes in
the random walk are picked from the concatenation of the neighbourhoods of all nodes previously selected to be a part of the
sampled subgraph. The authors present an unbiased estimator and compare it against \citeAutRef{chen2016general} and
\citeAutRef{han2016waddling}, getting better results than both for the single network tested.

\subsection{Colour Coding}

The technique of colour coding~\cite{alon1995color} has been adapted to the problem of approximating subgraph frequencies
by \citeAutRef{zhao2010subgraph}, \citeAutRef{zhao2012sahad} and \citeAutRef{slota2013fast}. However, all these
works focus on specific categories of subgraphs, for example, \textsc{SAHad}~\cite{zhao2012sahad} attempts to only find
subgraphs that are in the form of a tree. 

More recently, \citeAutRef{bressan2018motif} present a general algorithm using colour coding, that works for any undirected subgraph
of size theoretically unbound. The algorithm works in two phases. The first, based on the original description of~\cite{alon1995color},
is counting the number of non-induced trees, \emph{treelets}, in the graph but with a particularity, the nodes were previously
partitioned into $k$ sets and attributed a label (a \emph{colour}). These treelets then must be constituted solely of nodes with
different colours. This part of the algorithm outputs counters $C(T,S,v)$, for every $v \in V(G)$, which are the number of treelets
rooted in $v$ isomorphic to $T$, whose colours span the colour set $S$.

The second phase of the algorithm is the sampling part, which is focused on sampling treelets uniformly at random. To pick a treelet
with $k$ nodes, the authors choose a random node $v$, a treelet $T$ with probability proportional to $C(T,[k],v)$ and then pick
one of the treelets that is rooted in $v$, is isomorphic to $T$ and is coloured by $[k]$. Given a treelet $T_k$, the authors consider
the graphlet $G_k$ induced by the nodes of $T_k$ and increment its frequency by $\frac{1}{\sigma(G_k)}$, where $\sigma(G_k)$ is
the number of spanning trees of $G_k$.

\newcommand{\thickhline}{%
	\noalign {\ifnum 0=`}\fi \hrule height 1pt
	\futurelet \reserved@a \@xhline
}

\section{Parallel Strategies}\label{sec:parallel}

By this point it should be clear that subgraph counting is a computationally hard problem. As discussed in Section~\ref{sec:exact}, analytic approaches are much more efficient than enumeration algorithms; however, they are specific to certain sets of small subgraphs. Sampling strategies can produce results in a fraction of the time; but there's a trade-off between time and accuracy. Therefore, speeding up subgraph counting remains a crucial task. The availability of parallel environments, such as multicores, hybrid clusters, and GPUs gave rise to strategies that leverage on these resources. Here we follow a different organizational approach than Sections~\ref{sec:exact} and~\ref{sec:sampling}: we first give an historic overview of the parallel algorithms put forward throughout the years and then we discuss the strategies on a higher level. This is done because most parallel algorithms have a sequential counterpart (already described in previous sections) and many common aspects can be found between the parallel strategies. Table~\ref{tab:paralgs1} summarizes our proposed taxonomy composed 
of seven aspects, ordered by their publication year: (i) their computational \textbf{platform}, (ii) the \textbf{initial work-units} (what part of the graph is divided initially), (iii) the \textbf{runtime work-units} (what part of the graph is divided during runtime), (iv) the \textbf{search traversal} strategy (how the graph is explored), (v) the \textbf{work division} strategy (how work-units are distributed), (vi) how \textbf{work sharing} is performed (if applicable) between workers (e.g., CPU processors, or CPU/GPU threads), and (vii) if code is \textbf{publicly available}.

\begin{table}[!h]
	\footnotesize
	\centering
	\def\arraystretch{1.0}
	\caption{Parallel algorithms for subgraph counting.}
	\label{tab:paralgs1}
	\begin{tabular}{c^c^c^c^c^c^c^c^c^l }
		\rowstyle{\bfseries}
		& \multirow{2}{*}{Year} & \multirow{2}{*}{Platform} & \multicolumn{2}{c}{\bf Work-units} & Search & Work & Work & Public\\
		
		\rowstyle{\bfseries}
		& &  & Initial & Runtime & Traversal & Division & Sharing & Code \\[2pt] \hline
		& & & & & & & & \\[-6pt] 
		\rowstyle{}
		\textsc{ParWang}~\cite{wang2005parallel} & 2005 & DM & Vertices & \xmark & DFS & Static & \xmark & \xmark \\[2pt]
		\textsc{DM-Grochow}~\cite{schatz2008parallel}	& 2008  & DM & Isoclasses & Isoclasses & DFS  & First-Fit & \xmark & \xmark \\[2pt]
		\textsc{MPRF}~\cite{liu2009mapreduce}	& 2009 & MapReduce  & Edges & Subgraphs & BFS  & Static & \xmark & \xmark  \\
		\textsc{DM-ESU}~\cite{ribeiro2010parallelesu} & 2010 & DM   & Vertices & Subgraph-trees & DFS & Diagonal & M-W & \xmark \\
		\textsc{DM-Gtries}~\cite{ribeiro2010efficient} & 2010 & DM  & Vertices & Subgraph-trees  & DFS & Diagonal & W-W & \xmark  \\
		
		\textsc{SM-Gtries}~\cite{aparicio2014parallel} & 2014 & SM  & Vertices & Subgraph-trees & DFS & Diagonal   & W-W &\cite{gtscanner}\\
		\textsc{SM-FaSE}~\cite{aparicio2014scalable} & 2014 & SM & Vertices & Subgraph-trees & DFS &  Diagonal & W-W & \cite{gtscanner}\\
		\textsc{Subenum}~\cite{shahrivari2015fast} & 2015 & SM & Edges & Subgraphs & DFS & First-Fit  & \xmark & \cite{subenum}\\
		\textsc{GPU-Orca}~\cite{milinkovic14contribution} & 2015 & GPU & Vertices &  Subgraphs & BFS & Static & \xmark & \xmark\\
		\textsc{Lin}~\cite{lin2015network} & 2015 & GPU & Vertices & Subgraphs & BFS & Static & \xmark & \xmark\\ 
		\textsc{MRSUB}~\cite{shahrivari2015distributed} & 2015 & MapReduce & Edges & Subgraphs & BFS & Static & \xmark & \xmark\\
		\textsc{PGD}~\cite{ahmed2015efficient} & 2015 & SM  & Edges& \xmark & DFS & Static & \xmark & \cite{PGD}\\
		\textsc{GPU-PGD}~\cite{rossi2016leveraging} & 2016 & CPU+GPU & Edges & Subgraph-trees & BFS & First-Fit & W-W &  \xmark \\
		\textsc{Elenberg}~\cite{elenberg2016distributed} & 2016 & DM & Vertices & Subgraphs & DFS & First-Fit & \xmark & \cite{eelenberggit}\\
		\textsc{MR-Gtries}~\cite{ahmad2017scalable} & 2017 & MapReduce & Vertices & Subgraph-trees & DFS & Timed & M-W & \xmark\\
	\end{tabular}
\end{table}

\subsection{Historical Overview}\label{sec:par_hist}

One key aspect necessary to achieve a scalable parallel computation is finding a balanced work division (i.e., splitting work-units \emph{evenly} between workers -- parallel processors/threads). A naive possibility for subgraph counting is to assign $\frac{|V(G)|}{|P|}$ nodes from network $G$ to each worker $p \in P$. This egalitarian division is a poor choice since two nodes induce very different search spaces%
; for instance, $hub$-like nodes induce many more subgraph occurrences than nearly-isolated nodes. Instead of performing an egalitarian division, \citeAutRef{wang2005parallel} discriminate nodes by their degree and distribute them among workers, the idea being that each worker gets roughly the same amount of \textit{hard} and \textit{easy} work-units. Despite achieving a more balanced division than the naive version, there is still no guarantee that the node-degree is sufficient to determine the actual complexity of the work-unit. Distributing work immediately (without runtime adjustments) is called a \textbf{static division}. Wang et al. did not assess scalability in~\cite{wang2005parallel}, but they showed that their parallel algorithm was faster than \textsc{Mfinder}~\cite{milo2002network} in an E. Coli transcriptional regulation network. Since their method was not named, we refer to it as \textsc{ParWang} henceforth. %

\iffalse
The first approach targeting \underline{multicore machines} was by \textbf{Schreiber and Schw\"obbermeyer (FPF/MAVisto)} \cite{schreiber2005frequency,schreiber2005mavisto} and, similar to Wang et al.~\cite{wang2005parallel}, it was a static division method. They reported \textcolor{red}{speed-ups of $\approx$4-5x using a 8-core machine on ten different metabolic which were more significant when bigger subgraphs were counted (5 or 6 nodes)}.
Both \cite{wang2005parallel} and \cite{schreiber2005frequency} parallelized \underline{network-centric} sequential algorithms.\fi 

 The first parallel strategy with a \textbf{single-subgraph-search} algorithm at its core, namely \textsc{Grochow}~\cite{grochow2007network}, was by \citeAutRef{schatz2008parallel}. Since the algorithm was not named, and it targets a \textbf{distributed memory (DM)} architecture (i.e., parallel cluster), we refer to it as \textsc{DM-Grochow}. In order to distribute query subgraphs (also called \textbf{isoclasses}) among workers they employed two strategies: naive and \textbf{first-fit}. The naive strategy is similar to \textsc{ParWang}'s. In the first-fit model, each slave processor requests a subgraph type (or \textbf{isoclass}) from the master and enumerates all occurrences of that type (e.g., cliques, stars, chains). This division is \textbf{dynamic}, as opposed to static, but it is not balanced since different isoclasses induce very different search trees. For instance, in sparse networks $k$-cliques are faster to compute than $k$-chains. Using 64 cores, Schatz et al. obtained $\approx$10-15x speedups over the sequential version on a yeast PPI network. They also tried another novel approach by partitioning the network instead of partitioning the subgraph-set. However, finding adequate partitions for subgraph counting is a very hard problem due to partition overlaps and subgraphs traversing different partitions, and no speedup was obtained using this strategy. We should note that parallel graph partitioning remains an active research problem to this day~\cite{bulucc2016recent,meyerhenke2017parallel}, but is out of the scope of this work.
 
 All parallel algorithms mentioned so far traverse occurrences in a \textbf{depth-first (DFS)} fashion, since doing so avoids having to store intermediate states. By contrast, \citeAutRef{liu2009mapreduce} use a \textbf{breadth-first search (BFS)} where, at each step, all subgraph occurrences found in the previous one are expanded by one node. Their algorithm, \textsc{MPRF}, is implemented following a \textbf{MapReduce} model~\cite{dean2008mapreduce} which is intrinsically a BFS-like framework. In MPRF, mappers extend size $k$ occurrences to size $k+1$ and reducers remove repeated occurrences. At each BFS-level, \textsc{MPRF} divides work-units evenly among workers. We still consider this to be a static division since no adjustments are made in runtime. Thus, in our terminology, static divisions can be performed only once (at the start of computation in DFS-like algorithms) or multiple times (once per level in BFS-like algorithms). Overhead caused by reading and writing to files reduces \textsc{MRPF}'s efficiency, but the authors report speedups of $\approx7x$ on a 48-node cluster, when compared to the execution on a single-processor. 

 DFS-based algorithms discussed so far either perform a complete work-division right at the beginning (\textsc{ParWang}), or they perform a partial work-division at the beginning and then workers request work when idle (\textsc{DM-Grochow}). In both cases, a worker has to finish a work-unit before proceeding a new one. Therefore, it is possible that a worker gets stuck processing a very computationally heavy work-unit while all the others are idle. This has to do with work-unit granularity: work-units at the top of the DFS search space have high (coarse) granularity since the algorithm has to explore a large search space. BFS-based algorithms mitigate this problem because work-units are much more fine grained (usually a worker only extends his work-unit(s) by one node). The work by \citeAutRef{ribeiro2010parallelesu} was the first to implement \textbf{work sharing} during parallel subgraph counting, alleviating the problem of coarse work-unit granularity of DFS-based subgraph counting algorithms. Workers have a splitting threshold that dictates how likely it is to, instead of fully processing a work-unit, putting part of it in a global work queue. A work-unit is divided using \textbf{diagonal work splitting} which gathers unprocessed nodes at level $k$ (i.e., nodes that are reached by expanding the current work-unit) and recursively goes up in the search tree, also gathering unprocessed nodes of level $k-i$, $i < k$, until reaching level $1$. This process results in a set of finer-grained work-units that induces a more balanced search space than static and first-fit divisions. In \cite{ribeiro2010parallelesu} Ribeiro et al. use \textsc{ESU} as their core enumeration algorithm and propose a
 \textbf{master-worker (M-W)} architecture where a master-node manages a work-queue and distributes its work-units among slave workers. This strategy, \textsc{DM-ESU}, was the first to achieve near-linear speedups ($\approx$128x on a 128-node cluster) on a set of heterogeneous network. A subsequent version~\cite{ribeiro2010efficient} used \textsc{GTries} as their base algorithm and implemented a \textbf{worker-worker (W-W)} architecture where workers perform work stealing. \textsc{DM-Gtries} improves upon \textsc{DM-ESU }by using a faster enumeration algorithm (\textsc{GTries}) and having all workers perform subgraph enumeration (without wasting a node in work queue management). Similar implementations (based on W-W sharing and diagonal splitting) of \textsc{GTries} and \textsc{FASE}  were also developed for \textbf{shared memory (SM) environments}, which achieved near-linear speedups in a 64-core machine~\cite{aparicio2014parallel,aparicio2014scalable}. The main advantages of SM implementations is that work sharing is faster (since no message passing is necessary) and SM architectures (such as multicores) are a commodity while DM architectures (such as a cluster) are not.
 
 Instead of developing efficient work sharing strategies, \citeAutRef{shahrivari2015fast} try to avoid the unbalanced computation induced by vertice-based work-unit division. \textsc{Subenum} is an adaptation of \textsc{ESU} which uses edges as starting work-units, achieving near-linear speedup ($\approx$10x on a 12-core machine). Using edges as starting work-units is also more suitable for the MapReduce model since edges are finer-grained work-units than vertices. In a follow-up work~\cite{shahrivari2015distributed}, Shahrivari and Jalili propose a MapReduce algorithm, \textsc{MRSUB}, which greatly improves upon~\cite{liu2009mapreduce}, reporting a speedup of $\approx34$x on a 40-core machine. Like \textsc{Subenum}, \textsc{MRSUB} does not support work sharing between workers. A MapReduce algorithm with work sharing was put forward by \citeAutRef{ahmad2017scalable}, henceforth called \textsc{MR-Gtries}. Using work sharing with \textbf{timed redistribution} (i.e., after a certain time, every worker stops and work is fully redistributed), they report a speedup of $\approx26$x on a 32-core machine. While \textsc{MRSUB} and \textsc{MR-GTries} efficiency is comparable ($\approx80\%$), the latter has a much faster sequential algorithm at its core; therefore, in terms of absolute runtime, \textsc{MR-Gtries} is the fastest MapReduce subgraph counting algorithm that we know of.
 
 Graphics processing units (\textbf{GPUs}) are processors specialized in image generation, but numerous general purpose tasks have been adapted to them~\cite{fang2008parallel,hong2011efficient,merrill2012scalable}. GPUs are appealing due to their large number of cores, reaching hundreds or thousands of parallel threads whereas commodity multicores typically have no more than a dozen. However, algorithms that rely on graph traversal are not best suited for the GPU framework due to branching code, non-coalesced memory accesses and  coarse work-unit granularity~\cite{merrill2012scalable}. \textbf{Milinkovi\'c et al.}~\cite{milinkovic14contribution} were one of the firsts to follow a GPU approach (\textsc{GPU-Orca}), with limited success. \citeAutRef{lin2015network} put forward a GPU algorithm (henceforth refereed to as \textsc{Lin} since it was unnamed) mostly targeted at network motif discovery but also with some emphasis on efficient subgraph enumeration. \textsc{Lin} avoids duplicate in a similar fashion to ESU \cite{wernicke2005faster} and auxiliary arrays are used to mitigate uncoalesced memory accesses. A BFS-style traversal is used (extending each subgraph 1 node at a time) to better balance work-units among threads. They compare \textsc{Lin} running on a 2496-core GPU (Tesla K20) against parallel CPU algorithms and report a speedup of $\approx$10x to a 6-core execution of the fastest CPU algorithm, \textsc{DM-GTries}. 
 
 \citeAut{rossi2016leveraging} proposed the first algorithm that \textbf{combines multiple GPUs and CPUs}~\cite{rossi2016leveraging}. Their method dynamically distributes work between CPUs and GPUs, where unbalanced computation is given to the CPU whereas GPUs compute the more regular work-units. Since their method was not named, we refer to it as \textsc{GPU-PGD}. Their hybrid CPU-GPU version achieves speedups of $\approx 20$x to $\approx 200$x when compared to sequential \textsc{PGD}, depending largely on the network. As mentioned in Section~\ref{sec:exact}, \textsc{PGD} is one of the fastest methods for sequential subgraph counting. As such, \textsc{GPU-PGD} is the fastest subgraph counting algorithm currently available as far as we know. However, \textsc{GPU-PGD} is limited to 4-node subgraphs, while \textsc{DM-GTries} is the fastest general approach.

\subsection{Platform}

Different parallel platforms offer distinct advantages and are more suited for particular strategies. Next we discuss the strategic differences between platforms.

\subsubsection{Distributed Memory (DM)}

A parallel cluster offers the opportunity to use multiple (heterogenous) machines to speedup computation. Clusters can have hundreds of processors and therefore, if speedup is linear, computation time is reduced from weeks to just a few hours. For work sharing to be efficiently performed on DM architectures one can either have a master-node mediating work sharing~\cite{ribeiro2010parallelesu} or have workers directly steal work from each other~\cite{ribeiro2010efficient,ribeiro2012parallel}. Usually DM approaches are implemented directly using MPI~\cite{wang2005parallel,schatz2008parallel,ribeiro2010efficient,ribeiro2010parallel,ribeiro2012parallel} but higher level software, such as GraphLab, can also be used~\cite{elenberg2016distributed}. DM has the drawback of workers having to send messages through the network, making network bandwidth a bottleneck.

\subsubsection{Shared Memory (SM)}

SM approaches have the advantage in their underlying hardware being a commodity (multicore computers). Furthermore, workers in a SM environment do not communicate via network messages (since they can communicate directly in main memory), thus avoiding a bottleneck in the network bandwidth. However, the number of cores is usually very low when compared to DM, MapReduce, and GPU architectures. Algorithms on multicores tend to traverse the search space in a DFS fashion~\cite{aparicio2014parallel,aparicio2014scalable,shahrivari2015fast,ahmed2015efficient} thus avoiding the storage of large number of subgraph occurrences in disk or main memory.

\subsubsection{MapReduce}

The MapReduce paradigm has been  successfully applied to problems where each worker executes very similar tasks, which is the case of subgraph counting. MapReduce is an inherently BFS method, whereas most subgraph counting algorithms are DFS-based. The biggest setback of using MapReduce is the huge amount of subgraph occurrences that are stored in files between each BFS-level iteration (corresponding to a node expansion)~\cite{liu2009mapreduce,shahrivari2015distributed}. To avoid this setback, one can instead store them in RAM when the number of occurrences fits in memory~\cite{ahmad2017scalable}.

\subsubsection{GPU}

GPUs are very appealing due to their large amount of parallel threads. Despite linear speedups being rare in the GPU, since they have such a large number of cores the gains can still be substantial. However, they are not well-suited for graph traversal algorithms. One of current best pure BFS algorithms~\cite{merrill2012scalable} on the GPU only achieve a speedup of $\approx 8$x (on a 448-core NVIDIA C2050) when compared to a 4-core CPU BFS algorithm~\cite{leiserson2010work}. By contrast, Monte Carlo calculations on a NVIDIA C2050 GPU achieve a speedup of $\approx 30$x~\cite{ding2011evaluation} when compared to a 4-core CPU implementation. This is mainly due to branching problems, uncoalesced memory accesses and coarse work-unit granularity, sometimes leading to almost non-existent speedups in subgraph counting~\cite{milinkovic14contribution}. Using additional memory to efficiently store neighbors and smart work division help achieve some speedup~\cite{lin2015network}. Another approach is to combine CPUs and GPUs: CPUs handle unbalanced computation while GPUs execute regular computation~\cite{rossi2016leveraging}.

\subsection{Work-units}

\begin{figure}
	\centering
	\includegraphics[width=0.85\linewidth]{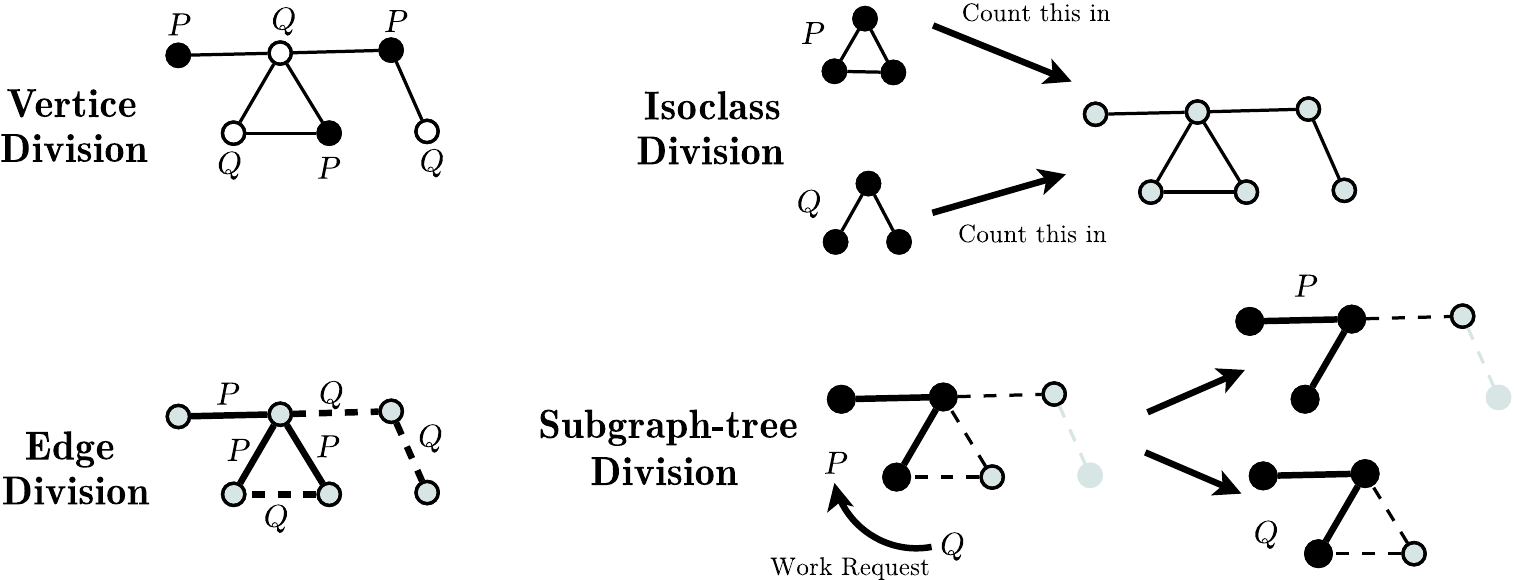}
	\vspace{-0.2cm}
	\caption{Different base work-units, and their division to two workers $P$ and $Q$.}
	\label{fig:workunits}
\end{figure}

When a program can be split into a series of (nearly) independent tasks, efficient parallelism greatly reduces execution times. Each \emph{worker} (be it a thread, CPU-core, or GPU-core) is assigned \emph{work-units} (parallel tasks), either at the start of the computation (\emph{initial work-units}) or during runtime (\emph{runtime work-units}). Work-units in subgraph counting can be either (a) \emph{vertices}, (b) \emph{edges}, (c) \emph{subgraphs}, (d) isomorphic classes (or \emph{isoclasses} for short), or (e) \emph{subgraph-trees}, and each option is described next. Figure~\ref{fig:workunits} illustrates each base work-unit and gives an example of how each can be divided.

\subsubsection{Vertices}\label{sec:wu_vertices} One possibility is to consider each vertex $v \in V(G)$ as a work-unit and split them among workers. A worker $p$ then computes all size-$k$ subgraph occurrences that contain vertex $v$. Naive approaches have different workers finding repeated occurrences that need to be removed~\cite{wang2005parallel}, but efficient sequential algorithms have canonical representations that eliminate this problem~\cite{wernicke2005faster,kashani2009kavosh,ribeiro2014g}, making each work-unit independent.  Using vertices as work-units has the drawback of creating very coarse work-units: different vertices induce search spaces with very different computational costs. For instance, counting all the subgraph occurrences that start (or eventually reach) a hub-like node is much more time-consuming than counting occurrences of a nearly isolated node. For algorithms with vertices as work-units to be efficient they can either try to find a good initial division \cite{wang2005parallel} or enable work sharing between workers \cite{ribeiro2010parallel,ribeiro2012parallel,aparicio2014parallel,aparicio2014scalable}. Each of these work division strategies is discussed in Section~\ref{sec:workdivision}.

\subsubsection{Edges}

Due to the unbalanced search tree induced by vertex division, some algorithms use edges as work-units \cite{liu2009mapreduce,shahrivari2015fast,ahmed2015efficient,shahrivari2015distributed}. The idea is similar to vertice division: distribute all $e(v_i, v_j) \in E(G)$ evenly among the workers.  An initial edge division guarantees that all workers have an equal amount of 2-node subgraphs, which is not true for vertex division. However, for $k \geq 3$ this strategy offers no guarantees in terms of workload balancing. Therefore, in regular networks (i.e. networks where all nodes have similar clustering coefficients) this strategy achieves a good speedup, but it is not scalable in general. Some methods~\cite{schatz2008parallel,rossi2016leveraging} perform dynamic first-fit division (discussed in Section~\ref{sec:firstfit}) instead the simple static division described.

\subsubsection{Subgraphs} At the start of computation, only vertices and edges from the network are known. As the $k$-subgraph counting process proceeds, subgraphs of sizes $k-i, i < k$ are found. Thus, the work-units divided among threads can be these intermediate states (incomplete subgraphs). Some BFS-based algorithms~\cite{liu2009mapreduce,milinkovic14contribution,shahrivari2015distributed,lin2015network} begin with either edges or vertices as initial work-units and, at the end of each BFS-level, intermediate subgraphs are found and divided among workers. DFS-based methods expand each subgraph work-unit by one node until they reach a $k$-subgraph~\cite{shahrivari2015distributed, elenberg2016distributed}.

\subsubsection{Isoclass} Instead of partitioning the graph, like the previous three strategies do, one can instead chose to partition the set of isoclasses being enumerated~\cite{schatz2008parallel}. Work-units split in this fashion have similar problems to the previously discussed: isomorphic classes do not induce computationally equivalent search spaces. For instance, in sparse networks it is much more time-consuming to enumerate chains or hubs than cliques. 

\subsubsection{Subgraph-trees} This approach is applicable only for DFS-like algorithms where, since the search tree is explored in a depth-first fashion, a work-tree is implicitly built during enumeration: when the algorithm is at level $k$ of the search, unexplored candidates of stages $\{k-1, k-2, ... , 1\}$ were previously generated. Then, instead of splitting top vertices from stage 1 only (as described in Section~\ref{sec:wu_vertices}), the search-tree is split among sharing processors \cite{ribeiro2010efficient,ribeiro2010parallel,aparicio2014parallel,aparicio2014scalable} (more details on this in Section~\ref{sec:diagonalsplit}). Subgraph-trees are \emph{expected} to be similar since both coarse- and fine-grained work-units are generated. Nevertheless, it is not guaranteed that work-units from the same level of the search tree induce similar work. This strategy also incurs the additional complexity of building the candidate-set of each level and splitting them among workers.

\subsection{Search Traversal}

Discounting analytic approaches presented in Section~\ref{sec:anal}, subgraph counting algorithms typically count occurrences by traversing the graph. How graph traversal is performed greatly influences the parallel performance and is dependent on the platform, as is discussed next.

\subsubsection{Breadth-First Search}\label{sec:bfs}

Algorithms that adopt this strategy are typically MapReduce methods~\cite{liu2009mapreduce,shahrivari2015distributed,ahmad2017scalable} or GPU~\cite{milinkovic14contribution,lin2015network,rossi2016leveraging} approaches. MapReduce works intrinsically in BFS fashion, and GPUs are very inefficient when work is unbalanced and contains branching code. BFS starts by (i) splitting edges among workers, (ii) the processors compute the patterns of size-3 from each edge (size-2 subgraphs), (iii) the patterns of size-3 are themselves split among processors and (iv) this process is repeated until the desired size-$k$ patterns are obtained. The idea of BFS is to give large amounts of fine-grained work-units to each worker, thus making work division more balanced since these work-units induce similar work, making this approach more suitable for methods that require regular data. However, the main drawback is that these algorithms need to store partial results (which grow exponentially as $k$ increases) and synchronize at the end of each BFS-level.

\subsubsection{Depth-First Search}

To avoid the cost of synchronization and of storing partial results, most subgraph counting algorithms traverse the search space in a depth-first fashion \cite{wang2005parallel,grochow2007network,ribeiro2010efficient,ribeiro2010parallel,ribeiro2012parallel,aparicio2014parallel,aparicio2014scalable,shahrivari2015fast,ahmed2015efficient}. This means that the algorithm starts with $V_{sub} = \{v\}$ and incrementally adds a new node to $V_{sub}$ until it obtains a match of the desired size, and backtracks to find new matches. This strategy leads to unbalanced search spaces, caused by coarse-grained work-units, that need to be controlled.

\subsection{Work Division}\label{sec:workdivision}

Splitting work is obviously essential for a parallel approach. Work can be divided at two moments: (i) an initial work division before subgraph counting starts and/or (ii) divisions during runtime. 

\subsubsection{Static} The simplest form of work division is to produce an initial distribution of work-units and proceed with the parallel computation, without ever spending time dividing work during runtime. Trying to obtain an estimation of the work beforehand~\cite{wang2005parallel,rossi2016leveraging} is valuable but limited: if the estimation is done quickly but is not very precise (such as using node-degrees or clustering coefficients to estimate work-unit difficulty) little guarantees are offered that the work division is balanced, and obtaining a very precise estimation is as computationally expensive as doing subgraph enumeration itself. Following a BFS approach~\cite{liu2009mapreduce,milinkovic14contribution,shahrivari2015distributed} helps balancing out the work-units and a static work division at each BFS-level is usually sufficient to obtain good results. However, those strategies have limitations as discussed in Section~\ref{sec:bfs}. Some analytic works, which do not rely on explicit subgraph enumeration, do not need advanced work division strategies because their algorithm is almost embarrassingly parallel~\cite{rossi2017estimation}. 

\subsubsection{Dynamic: First-fit}\label{sec:firstfit} Instead of trying to estimate a good division one can generate work on-demand during runtime. One isomorphic class~\cite{schatz2008parallel} or small portions of the graph~\cite{shahrivari2015fast} can be initially given at each processor and, when that computation is done, idle processors request more work. This strategy has the penalty of maintaining a global queue of work-units to be processed. Furthermore, the last $|P|$ work-units (where $|P|$ is the number of workers) can have different granularities (and thus computational cost), so the speedup is largely dependent on how well-balanced they are.

\subsubsection{Dynamic: Diagonal Work Splitting}\label{sec:diagonalsplit} Algorithms that employ this strategy~\cite{ribeiro2010efficient,ribeiro2010parallel,aparicio2014parallel,aparicio2014scalable} perform an initial static work division. They do not need a sophisticated criteria to choose to whom work-units are assigned because work will be dynamically redistributed during runtime: whenever workers are idle, some work will be relocated from busy workers to them. Furthermore, instead of simply giving half of their top-level work-units away and keeping the other half, a busy worker fully splits its work tree
The main idea is to build work-units of both fine- and coarse-grained sizes,  and this is particularly helpful in cases where a worker becomes stuck managing a very complex initial work-unit; this way, that work-unit is split in half, and it can be split iteratively to other workers if needed. These work-units can then either be stored in a global work queue, which a master worker is responsible of managing~\cite{ribeiro2010efficient,ribeiro2010parallel}, or sharing is conducted between worker threads themselves~\cite{aparicio2014parallel,aparicio2014scalable} (more details on Sections~\ref{sec:masterworker} and~~\ref{sec:workerworker}, respectively).

\subsubsection{Dynamic: Timed Redistribution}\label{sec:adaptiveredist} Timed Redistribution is a way to avoid estimating work during runtime while guaranteeing that every worker has work (after a while). Workers first receive work and try to process as much as they can. After a certain time, they all stop and work is redistributed. This strategy is specially useful when worker communication is not practical, such as in a MapReduce environment~\cite{ahmad2017scalable} on in the GPU. Setting an adequate threshold for work redistribution has a great impact: redistributing work too quickly has the drawback of wasting too much time in work division, while redistributing work too late has the drawback of having idle workers. One solution is to use an adaptive threshold~\cite{ahmad2017scalable}: if workers are too often without work, the threshold of the next iteration is lower, if workers are too often with much work left to compute, the threshold of the next iteration is higher.

\subsection{Work Sharing}

Since work is unbalanced for enumeration algorithms, work sharing can be used in order to balance work during runtime. 

\subsubsection{Master-Worker (M-W)}\label{sec:masterworker}

This type of work sharing is mostly adopted in distributed memory (DM) environments since workers do not share positions of memory that they can easily access and use to communicate. A master worker initially splits the work-units among the workers (slaves) and then manages load balancing. Load balancing can be achieved by managing a global queue where slaves put some of their work, to be later redistributed by the master~\cite{ribeiro2010parallelesu}. This strategy implies that the master is not being used the enumeration and that there is a need communication over the network.

\subsubsection{Worker-Worker (W-W)}\label{sec:workerworker} 

Shared memory (SM) environments allow for direct communication between workers, therefore a master node is redundant. In this strategy, an idle worker asks a random worker for work~\cite{aparicio2014parallel,aparicio2014scalable}. One could try to estimate which worker should be polled for work (which is computationally costly) but random polling has been established as an efficient heuristic for dynamic load balancing \cite{sanders1994detailed}. After the sharing process, computation resumes with each worker evolved in the exchange computing their part of the work. Computation ends when all workers are polling for work. This strategy achieves a balanced work-division during runtime, and the penalty caused by worker communication is negligible~\cite{aparicio2014parallel,aparicio2014scalable}. Most implementations of W-W sharing are built on top of relatively homogeneous systems, such as multiworkered CPUs~\cite{shahrivari2015fast} or clusters of similar processors~\cite{schatz2008parallel}. In these systems, since all workers are equivalent, it is irrelevant which ones get a specific easy (or hard) work-unit, thus only load balancing needs to be controlled. Strategies that combine CPUs with GPUs, for instance, can split tasks in a way that takes advantage of both architectures: GPUs are very fast for regular tasks while CPUs can deal with irregular ones. For instance, a shared deque can be kept where workers, either GPUs or CPUs, put work on or take work from~\cite{rossi2016leveraging}; the queue is ordered by complexity: complex tasks are placed at the front, and simple tasks at the end. The main idea is that CPUs handle just a few complex work-units from  the front of the deque while GPUs take large bundles of work-units from the back.

\section{Concluding remarks}\label{sec:conclusions}

Over the last twenty years, subgraph counting has been under increased focus in the network science community, specially since the introduction of networks motifs~\cite{milo2002network}, and its status as an important tool for network analysis, as well as graphlets~\cite{przulj2007} which are now established measures for network alignment.

In this survey we explored existing practical methods to solve the subgraph counting problem from three perspectives: (i) algorithms that efficiently perform exact counting, which is an intrinsically computationally hard task, (ii) algorithms that perform an approximation of subgraph frequencies, making the process faster but taking into account the accuracy of their estimation, and (iii) algorithms that efficiently exploit parallel architectures despite the unbalanced nature of subgraph counting. We showed that all three of these categories are still attracting new work, with new methods still emerging in an attempt to improve previous work.

The aim of this work was precisely to describe and classify the major algorithmic ideas in each of these three categories, offering a structured and thorough review of how they work and what are their advantages and disadvantages. Moreover, we provided more than two hundred references that allow further exploration of any aspects that might be of particular interest to the reader, including direct links to the existing practical implementations of the methods.

We feel that this survey provides valuable insight both from a more practical point of view, offering solutions and application ideas for those who view subgraph counting as a tool for network analysis, and from a more methodological angle, being not only a very strong starting point for new researchers joining the area, but also a very useful and comprehensive summary of recent research results for more established researchers.

\bibliographystyle{ACM-Reference-Format}
\bibliography{survey}
\end{document}